\documentclass[journal]{vgtc}                %
\ifpdf%
  \pdfoutput=1\relax                   %
  \usepackage{graphicx}                %
  \DeclareGraphicsExtensions{.pdf,.png,.jpg,.jpeg} %
\else%
  \usepackage{graphicx}                %
  \DeclareGraphicsExtensions{.eps}     %
\fi%

\graphicspath{{graphics/comparison/}{graphics/}} %

\usepackage{microtype}                 %
\PassOptionsToPackage{warn}{textcomp}  %
\usepackage{textcomp}                  %
\usepackage{times}                     %
\usepackage{cite}                      %
\usepackage{tabu}                      %
\usepackage{booktabs}                  %
\onlineid{1352}

\vgtccategory{Research}
\vgtcpapertype{system}

\usepackage[utf8]{inputenc}
\usepackage{amsmath,amssymb,color,graphicx}
\usepackage{tabularx,multirow}
\usepackage[usenames,dvipsnames,svgnames,table]{xcolor}
\usepackage{blkarray}
\usepackage{verbatim}
\usepackage{adjustbox}

\usepackage{xspace}
\usepackage{caption}
\usepackage{subcaption}
\usepackage{wrapfig}
\usepackage{rotating}

\usepackage{amsmath}
\usepackage{algorithm}
\usepackage[noend]{algpseudocode}
\usepackage{mathtools}
\usepackage{array}

\definecolor{Gray}{gray}{0.85}

\newcolumntype{a}{>{\columncolor{Gray}}l}

\usepackage[misc]{ifsym}  %
\usepackage[normalem]{ulem}

\usepackage{hyperref}

\usepackage{xcolor}

\usepackage{multirow}
\usepackage{multicol}

\usepackage{enumitem}

\newcommand{\hH}{\mathcal H}
\newcommand{\hV}{V}
\newcommand{\hE}{\mathcal S}

\newcolumntype{Y}{>{\centering\arraybackslash}X}

\definecolor{paleGreen}{RGB}{148, 248, 148}
\definecolor{medGreen}{RGB}{54, 186, 70}
\definecolor{deepGreen}{RGB}{12,134,30}

\newcommand{\pg}{\cellcolor{paleGreen}}
\newcommand{\mg}{\cellcolor{medGreen}}
\newcommand{\dg}{\cellcolor{deepGreen}}

\newcommand{\new}[1]{#1}

\makeatletter
\newcommand{\iflabelexists}[3]{\@ifundefined{r@#1}{#3}{#2}}
\makeatother

\teaser{
  \centering
  \includegraphics[width=\linewidth]{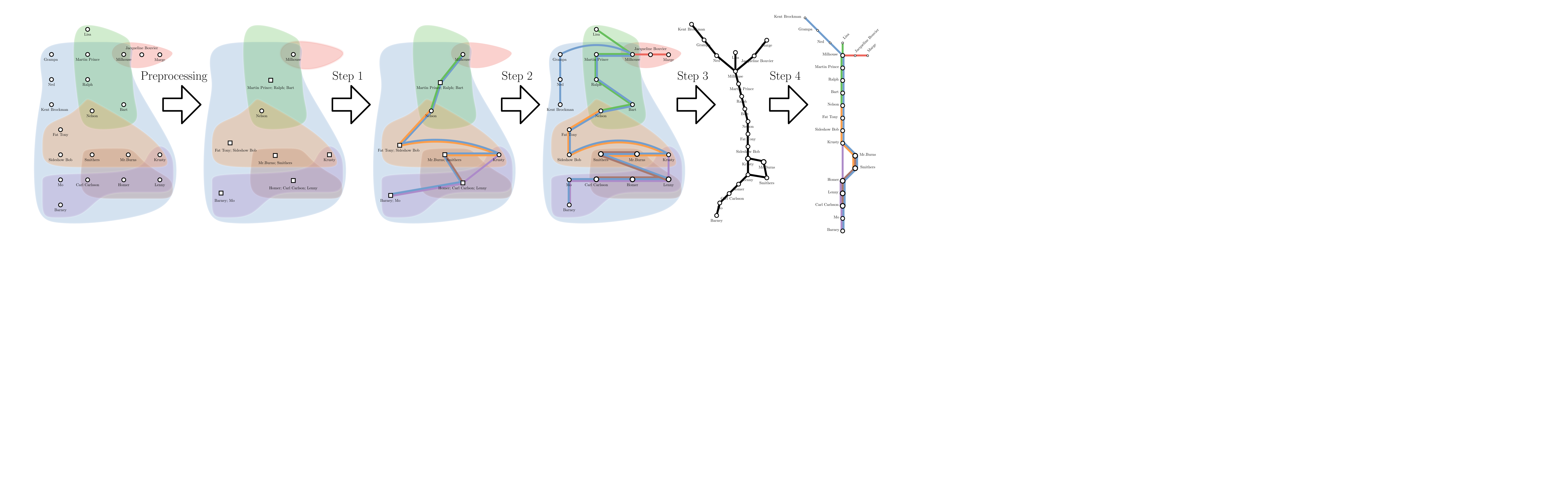}
  \caption{The MetroSets pipeline. The input set system (here: characters of \textit{The Simpsons}) is first compressed into a combinatorially equivalent smaller instance; Step 1 creates an optimized path support graph; Step 2 reinserts temporally discarded elements; Step 3 creates an initial layout of the support graph; finally Step 4 schematizes the layout as a metro map and places the labels. }
  \label{fig:teaser}
}

\vgtcinsertpkg


\title{MetroSets:  Visualizing Sets as Metro Maps}

\author{Ben Jacobsen, Markus Wallinger, Stephen Kobourov, and Martin N\"ollenburg}
\authorfooter{
\item
 Ben Jacobsen is with University of Arizona. E-mail: bjacobsen@email.arizona.edu
\item
 Markus Wallinger is with TU Wien. E-mail: mwallinger@ac.tuwien.ac.at
\item
 Stephen Kobourov is with University of Arizona. E-mail: kobourov@cs.arizona.edu
\item 
Martin N\"ollenburg is with TU Wien. E-mail: noellenburg@ac.tuwien.ac.at
}

\shortauthortitle{Jacobsen \MakeLowercase{\textit{et al.}}: MetroSets: Visualizing Sets as Metro Maps}
\renewcommand{\manuscriptnotetxt}{}

\abstract{We propose MetroSets, a new, flexible online tool for visualizing set systems using the metro map metaphor.
   We model a given set system as a hypergraph $\hH = (\hV, \hE)$, consisting of a set $\hV$ of vertices and a set $\hE$, which contains subsets of $\hV$ called hyperedges.
   Our system then computes a metro map representation of $\hH$, where each hyperedge $E$ in $\hE$ corresponds to a metro line and each vertex corresponds to a metro station.
   Vertices that appear in two or more hyperedges are drawn as interchanges in the metro map, connecting the different sets.
   MetroSets is based on a modular 4-step pipeline which constructs and optimizes a path-based hypergraph support, which is then drawn and schematized using metro map layout algorithms.
  We propose and implement multiple algorithms for each step of the MetroSet pipeline and provide a functional prototype with \new{easy-to-use preset configurations.} %
Furthermore, \new{using several real-world datasets}, we perform an extensive quantitative evaluation of the impact of different pipeline stages on desirable properties of the generated maps, such as octolinearity, monotonicity, and edge uniformity.
	}
 
\begin{document}

\firstsection{Introduction} 

\maketitle

We describe MetroSets: a flexible and modular system for visualizing set systems using the schematic metro map metaphor. The sets are represented by metro lines and set elements are represented by metro stations, with elements that belong to multiple sets corresponding to interchange stations; see \autoref{fig:teaser}. This approach leverages the familiarity of metro maps and scales well for medium-sized datasets with over a dozen sets and hundreds of elements.

Set systems arise in many applications, from astronomy to zoology, and analyzing their structure is a difficult problem; see the
survey
by Alsallakh et al.~\cite{alsallakh2014visualizing}. Such analysis can be aided greatly by effective data visualization. 
Perhaps the best known set visualizations are the Venn and Euler diagrams.
While being intuitive and easy to understand, they do not scale well and cannot be used to visualize more than a few sets at once. Visual analytics systems, such as UpSet~\cite{lex2014upset}, make it possible to work with much larger datasets. However, this scalability often comes at the cost of a non-trivial interface and less intuitive visualizations. \new{Our work focuses on datasets which fall in the middle: too large to be effectively visualized with Euler Diagrams, but not so large that they require a full-fledged visual analytics system.} 

The large number of hand-made metro map visualizations and infographics of set systems~\cite{antoniazzi,Booth,Honnorat,Simmonds,brexit19} is evidence of the appeal of this approach. Prior work on the effectiveness of map-like visualization is also encouraging~\cite{saket2014node,saket2015map,saket2016comparing}.  The difficulty is that creating such visualizations requires both expert knowledge and a considerable investment of time. With this in mind, we propose the MetroSets system: a modular online platform which automates the process, making it possible for everyone to efficiently  generate attractive metro map visualizations of set systems, while optimizing different desirable properties, such as minimal edge crossings, monotone metro lines, octolinear slopes, and uniform distances between adjacent stations.

Our design goals are matched to different optimizations in the 4-stage MetroSets pipeline. For example, representing all elements in the same pairs of sets as one contiguous curve is captured by optimizing consecutive ones during the creation of a planar support graph, while realizing uniform distances between adjacent metro stations is captured by optimizing edge uniformity in the layout of the planar support graph. Each of these properties can be evaluated via quantifiable measures and extensive experiments show that even though the underlying problems are NP-hard, our  heuristics perform well in practice. MetroSets provides several interactions, making it suitable for 15 of the set visualization tasks of Alsallakh et al.~\cite{alsallakh2014visualizing}. Finally, MetroSets scales well with respect to running time, visualizing set systems with over one hundred elements and more than a dozen sets in a matter of seconds, as shown by a large-scale evaluation using several thousand datasets. 
Please refer to the video in the supplemental materials to see MetroSets in action.

\section{Related Work}\label{se:relatedwork}

\subsection{Set Visualization}

While sets and set-typed data have not received as much attention in the visualization literature as other data types, sets, e.g., defined by groups with the same values of categorical attributes, are frequently included as secondary aspects in visualizations. 
In contrast, we focus our attention to sets and their relations as the primary aspect of the data.
In 2016, Alsallakh et al.~\cite{alsallakh2014visualizing} presented an in-depth state-of-the-art report on set visualization, which defines a taxonomy of element-related, set-related, and attribute-related tasks for set visualizations as well as a classification of set visualization techniques.

The oldest and most popular set visualizations are Euler diagrams, representing each set with a closed curve %
and set relations with curve intersections.
Euler diagrams generalize Venn diagrams, which show every possible relation of the given sets.
Most of the time, Euler diagrams do not explicitly show individual set elements, but some methods aim for area proportionality to indicate set cardinalities. 
Methods for automatically generating Euler diagrams use circles and ellipses for the sets~\cite{sfrh-adedwc-12,w-eaacved-12,larsson2020eulerr,mr-efled-14}, as well as less regular shapes~\cite{rd-ued-10,saa-favos-09,sas-sabied-16,rzf-gedg-08,srhz-iged-11}.
While Euler and Venn diagrams are intuitive, they have limited scalability and work well only for a handful of different sets~\cite{alsallakh2014visualizing}. 
Since well-formed Euler diagrams do not always exist, some of the algorithms may also produce inconsistent diagrams and non-existing set intersections.
Our aim with MetroSets is to develop an equally intuitive set visualization technique using the well-known metro map metaphor, but be able to better scale to larger numbers of sets with more complex overlaps and also explicitly show all individual elements of the sets.

Another type of set visualization is based on overlaying set contours on elements with pre-specified locations, e.g., spatial set data or sets defined on top of node-link network drawings. 
Here, the set aspects of the data are only secondary and the set visualization must adapt to the geometric input.
Bubble Sets~\cite{cpc-bsrrwioev-09} %
draw isocontours enclosing prespecified point sets based on the point locations, and GMap~\cite{hu2010visualizing} computes Vornoi-based regions based on clusters in an embedded graph. Both Bubble Sets and GMap may show non-existing set intersections. 
MapSets~\cite{ehkp-mvecg-15} also enclose sets by contour curves, but using a space-filling, convexity-optimizing, and non-overlapping style inspired by political maps. 
Line Sets~\cite{ahrc-dslnvt-11} is based on the idea of connecting the elements of each set by a smooth and short curve that is optimized with a traveling salesperson heuristic. 
There are certain visual similarities to the metro map metaphor such as connecting the elements of each set as a line or showing elements belonging to multiple sets as interchanges.
However, other aspects are quite different from metro maps, e.g., sequences of elements belonging to two or more sets are not shown as parallel lines and the shape of curves is not simplified and schematized, but rather irregular and they may have many crossings without semantic meaning.
The latter is due to the fixed positions of the set elements, which is also why Line Sets cannot be applied to abstract set systems without given element locations.
Kelp Diagrams~\cite{dksw-kdpmv-12} and KelpFusion~\cite{mhsad-khvt-13} extend and combine ideas from LineSets and Bubble Sets by defining a parameter space ranging from sparse spanning graphs to convex hulls, with the middle range resulting in bubble shapes for local point clusters in a set and thinner links to bridge longer distances.
All the overlay techniques assume/require pre-computed point positions and so these methods are not directly applicable for abstract set systems as the ones considered in this paper.

The next group are techniques based on bipartite graphs, where each set is represented as a vertex and each set element as a second type of vertex. 
Then each element is connected by an edge to all sets containing that element. 
Visualizations for bipartite graphs thus become of interest and have been integrated in several systems~\cite{sgl-jsiativ-08,m-amvtdbg-07,drrd-pstfis-12,be-dhss-00,aamh-rsivalos-13}.
However, for complex set systems, the resulting layouts are dense with many edge crossings and there is little support for set-related tasks~\cite{alsallakh2014visualizing}.

Hypergraph layout approaches are also relevant as a set system is naturally modeled by a hypergraph. Johnson and Pollak~\cite{jp-hpcdvd-87} introduced different notions of hypergraph planarity. Many papers studied support graphs (or supports in short), which are graphs defined on the elements as vertex set, such that the elements of each set (or hyperedge) induce a connected subgraph, which could then be used as the basis for a set visualization. 
Of particular interest are planar supports~\cite{bkmsv-psh-11a,cgmny-spssh-19}, tree supports~\cite{ks-cmoct-03,kmn-mtshled-14}, and path supports~\cite{bcps-psh-12}.
We note that not all hypergraphs admit all types of supports, which may limit their general-purpose use in practice.
In our first pipeline step, we actually compute a (not necessarily planar) path support.

Matrix-based approaches map sets and elements to rows and columns, respectively, and marking containment by a dot. 
This makes it easy to determine all elements of a set or all sets an element belongs to. 
However, more complex set-related tasks typically require interaction, so most systems are designed for interactive analysis and exploration.
Examples are ConSet~\cite{kls-vcwpmd-07}, OnSet~\cite{smds-ovtlbd-14}, RainBio~\cite{lamy2019rainbio}, or UpSet~\cite{lex2014upset}.
The latter is a powerful and scalable visual analytics system with many possibilities for interactive queries. 
Matrix-based methods scale well, but have a strong dependency on row and column ordering, are less intuitive, and require interaction with a non-trivial interface for more complex tasks.

In linear diagrams~\cite{rsc-vswld-15,lm-clmdv-19,stapleton2019efficacy} each set is a row in a table filled with one or more horizontal line segments. Any vertical line intersecting the diagram crosses a certain subset of horizontal line segments, which indicates that these sets have a non-empty intersection (similar to an overlap region in an Euler diagram). 
Linear diagrams typically do not show individual elements of the sets, but only the set relations.

\subsection{Metro Map Layout}\label{sub:mmlayout}

The area of schematic metro map generation is well researched and several papers have been published on this subject the last few years~\cite{nw-dlhqm-11,wc-fmm-11,cr-oflwmpsd-14,wp-ime-16,srmw-amlumo-11,lutz2014realtime, bast2020}. 
An overview of different algorithms up to 2014 is given in Nöllenburg's survey paper~\cite{nollenburg2014survey}. 
The recent state-of-the-art report  by Wu et al.~\cite{wntrn-stlfdmhp-20} discusses the latest works on metro map layout from a human, machine, and design perspective.
These two surveys cover most of the literature on schematic metro map generation.
For applying such a schematization algorithm as Step 4 of our pipeline, a geometric input layout is needed, which is usually derived from the physical position of rail tracks but, in the case of abstract graphs, from an intermediate graph layout (computed in Step 3 of the pipeline).

\subsection{Metro Map Metaphor}

The popularity and ubiquity of metro maps in large cities worldwide~\cite{o-mmw-03} has turned them them a natural metaphor for artists and graphic designers.
The familiarity and simplicity of well-designed metro maps make them an attractive choice for catching the attention of people and letting them delve into exploring the depicted information. 
In fact, visualizations of all kinds of data have been turned into metro-map-like pictures.
Some works explore the usefulness of the metro map metaphor by studying hand-drawn maps for various data sets. A few examples are metro maps of cancer pathways~\cite{hw-smcp-02}, cell signaling~\cite{makieva2018inside}, politics~\cite{brexit19}, music~\cite{antoniazzi}, and project plans~\cite{nesbitt2004getting}. 
Others provide an interactive editor with some layout support, but most map creation steps being made by the user~\cite{sandvad2001metro}.
Finally, there are also multiple implemented systems and algorithms for automatically creating visualizations inspired by the metro map metaphor.
The main challenge of these works is to turn the respective data into a graph, which in a second step is to be drawn as a metro map.

The first step is highly application dependent and includes tree-like neurite structures~\cite{al2014neurolines}, coherent themes in newspaper articles and other text documents~\cite{ShahafGH12,shahaf2015information}, plant disease progression~\cite{wahabzada2015metro}, air traffic routes~\cite{hurter2010automatic}, and project plans~\cite{aguirregoitia2010software,srbms-alppumm-05}.
In all these examples, the definition of metro lines is either directly derived from ordered data (e.g., using temporal attributes) or extracted through more complex data analysis~\cite{ShahafGH12,shahaf2015information}.
None of these tools can create metro maps of set systems, nor do they make use of hypergraph supports. Most of the examples reported in the literature have a low degree of interaction between the different metro lines.
The final layout component is created either by a simple force-based method or uses/adapts one of the existing metro map layout algorithms (Section~\ref{sub:mmlayout}).
For temporal data, time is usually mapped to the x-axis, such that the layout only needs to compute the y-coordinates.
Such temporal metro maps, where all lines are x-monotone, are quite similar to Storyline visualizations~\cite{foo,tm-dcosv-12,lwwll-stes-13}.

\section{The MetroSets Pipeline}\label{se:pipeline}

Throughout this paper, we model the input set system as a hypergraph $\hH = (\hV,\hE)$. Hypergraphs are a generalization of ordinary graphs which allow edges to contain any number of vertices. These hyperedges then correspond to sets, while their vertices correspond to elements.

A metro map layout of a hypergraph is a drawing of a graph $G=(V,E)$ with the property that every hyperedge $E \in \hE$ corresponds to a path $v_0, v_1,...,v_n$ in $G$. \new{This drawing should follow the conventions of the metro map metaphor, spelled out in the design goals below.} %

We describe a framework for creating such drawings from input hypergraphs. The framework is modeled as a flexible, four-step pipeline with mandatory pre- and postprocessing steps (see \autoref{fig:teaser}).

\begin{itemize}[noitemsep]
    \item In the \textbf{preprocessing} step we set aside and merge vertices from the input hypergraph to decrease the problem size. 
    \item \textbf{Step 1} constructs a path-based support that defines a linear order over the vertices of each hyperedge. 
    \item \textbf{Step 2}  expands the merged vertices and reintroduces the removed vertices from the preprocessing step.
    \item \textbf{Step 3} creates an initial embedding of the support graph. 
    \item \textbf{Step 4} schematizes the initial layout by straightening paths, standardizing edge lengths, and setting angles to multiples of $45^{\circ}$. %
    \item In the \textbf{postprocessing} step, we determine optimal placement for labels and ordering for lines along each edge.
\end{itemize}

 \new{We implemented two or more different methods for each of the pipeline steps. To simplify the interface, we additionally provide three preset pipeline configurations which are easy to use and suitable for most tasks.}

\subsection{Design Goals}\label{sec:designgoals}
The design goals guiding  MetroSets and its pipeline steps can be partitioned into (i) design goals for creating and optimizing the path support $G$ and (ii) design goals for optimizing the metro map layout of $G$.

\paragraph{Path Support.}

While real metro maps have the topology defined by geographical coordinates and physical connections between stations, we can freely order vertices along each metro line. We define the following design goals that are crucial for being able to find a good layout in pipeline Steps 3 and 4.   

\begin{itemize}[noitemsep]
    \item \textbf{Conjointness/Closeness.} Vertices that share the same set of hyperedges should be drawn closely together with conjoint lines that avoid branching off and rejoining.
    \item \textbf{Sparseness.} To aid in creating readable layouts, we prefer balanced support graphs that avoid excessively connected, dense centers.
    \item \textbf{Planarity.} The support graph should allow a planar embedding.
\end{itemize}

\paragraph{Metro Map Layout.}
One reason for the success of the metro map metaphor in general is that most metro maps worldwide follow a very similar set of design rules~\cite{o-mmw-03,nollenburg2014survey}. Therefore our goal for the metro map layout style implemented for MetroSets is to be as close to the appearance of real metro maps as possible. In particular we aim to achieve the following design goals.

\begin{itemize}[noitemsep]
    \item \textbf{Octolinearity.} All edges of the metro map should have an octolinear orientation.
    \item \textbf{Straightness/Monotonicity.} %
    Individual metro lines should have few bends and obtuse bend angles; they should run monotonically through the map.
    \item \textbf{Edge Crossings.} Crossings between metro lines that are not interchange stations should be avoided, especially if the support is planar.
    \item \textbf{Edge Uniformity.} The distances between pairs of adjacent stations should be as uniform as possible. 
    \item \textbf{Station Separation.} Unrelated metro lines and stations should keep a sufficiently large distance. 
    \item \textbf{Line Crossings.} Parallel metro lines sharing the same tracks should cross as little as possible.
    \item \textbf{Colors and Symbols.} Metro lines should have contrasting and distinguishable colors. Stations are shown as small circle symbols with larger circles for interchanges.
    \item \textbf{Label Placement.} Stations should be  labeled unambiguously by their names without overlaps, ideally horizontally aligned and coherently on the same side of each line.
\end{itemize}

 \new{Note that not all  design goals align well. For example,  monotonicity and octolinearity can literally and figuratively pull the layout in different directions. With this in mind, we opt for a modular multi-step pipeline with different algorithms for each stage, making it possible to add new methods or design goals. As having many options and choices could make MetroSets difficult to use, we provide several “presets:” balanced, simplicity, max-speed. Each of these presets selects a complete path through the pipeline, requiring no options to be selected or parameters to be set.}

\subsection{Preprocessing and Support Graph Extraction (Step 1)}

Because the support graph construction is computationally expensive, we begin by preprocessing the input hypergraph into a condensed version which preserves all set intersections. We do this by temporarily discarding all vertices which belong to only a single hyperedge, and merging together all vertices which belong to the same hyperedges. The discarded and merged vertices are then returned after the support graph has been constructed.

The input to the first pipeline step is then the compressed hypergraph $\hH = (\hV,\hE)$, where $\hV$ is the set of vertices, and $\hE \subseteq 2^\hV$ is the set of hyperedges. 
The desired output of Step 1 is a path-based support graph $G = (V,E)$, which has the property that every hyperedge $E \in \hE$ induces a Hamiltonian subgraph in $G$, i.e., the vertices of $E$ can be spanned by a path~\cite{bcps-psh-12}. 
To do so, we must determine an order in which this path should visit the vertices in each hyperedge. 
The union of these ordered paths, one for each hyperedge, is our support graph $G$. 
We propose two methods for constructing the support graph.

\subsubsection{Two-Opt Heuristic}

The motivation behind this method is the idea that vertices which belong to the same or similar groups of hyperedges should be placed closely together in our final drawing, to emphasize their similarity. Thus, we want to choose our paths in such a way that vertices which share many hyperedges are visited sequentially.
 To this end, we assign each pair of vertices along a hyperedge a similarity score, which is \new{the number of hyperedges that both belong to. For example, if vertex $u$ belongs to the hyperedges $\lbrace A,B,C \rbrace$ and vertex $v$ belongs to the hyperedges $\lbrace B,C,D \rbrace$, then their similarity score is $2$. We then treat the problem of finding an optimal order in which to visit the vertices as a travelling salesperson (TSP) path problem, where the cost of an edge between two vertices is the reciprocal of their similarity score.} We employ the two-opt heuristic\cite{croes1958method} to find such a path.

\begin{figure}[tb]
    \centering
    \includegraphics[width=0.9\columnwidth]{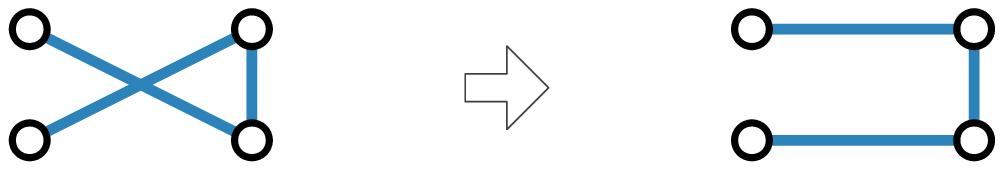}
    \caption{Illustration of a two-opt move, the local improvement used by the two-opt heuristic. Two edges are removed from the path and  vertices are reconnected to reduce the total length of the path.}
    \label{fig:2opt-illus}
\end{figure}

The two-opt heuristic is a local search method for solving TSP instances. While most TSP heuristics, including two-opt, are defined in terms of finding good cycles, the two-opt heuristic can be easily modified to find a good path. The two-opt heuristic also has the benefit of not requiring that the underlying space is metric.
Two-opt starts with an initial path, removes two edges, and then reconnects the dangling vertices to create a new path (called a two-opt move; see \autoref{fig:2opt-illus}). It then checks to see if the new path is shorter than the initial one. After attempting every possible two-opt move, it chooses the one which produces the largest improvement. This process is repeated until every possible move fails to improve upon the initial path, at which point we say the path is two-optimal~\cite{johnson1997traveling}.
We construct our initial path using the nearest neighbor heuristic, whereby we start at an initial vertex and recursively move to the closest neighbor that hasn not been visited yet. We then improve this path with the two-opt heursitic.

The two-opt algorithm is well studied, in part because it is a key component of the popular Lin-Kernighan algorithm\cite{lin1973effective}. Despite a relatively poor approximation ratio of $\sqrt{n/2}$, where $n$ is the number of cities to be visited, it often performs well in practice~\cite{hougardy2019approximation}. Because our TSP problem is non-metric and involves finding a path, rather than a cycle, we performed extensive quantitative analysis on random TSP instances to decide if the two-opt heuristic was appropriate for our use case. We found that, when using the nearest neighbor heuristic for initial route construction, the two-opt heuristic produced results that were on average only 5\% worse than the optimal solution, found using an ILP solver; more details in supplementary materials, section \ref{sec:eval2opt}%

\subsubsection{Consecutive Ones Method}

\begin{figure}[t]
  \centering
  \begin{subfigure}[b]{0.49\columnwidth}
    \centering
    \includegraphics[width=\linewidth]{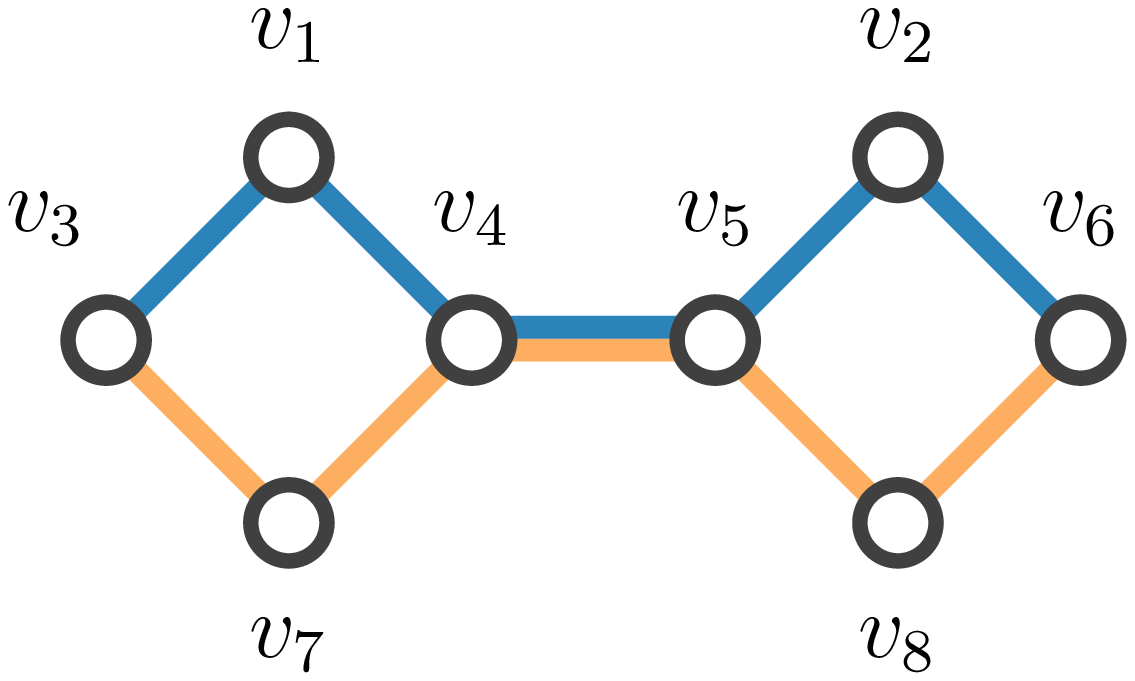}
    \caption{Path-based support of $\Pi_1$.}
  \end{subfigure}
  \begin{subfigure}[b]{0.49\columnwidth}
    \centering
    \includegraphics[width=\linewidth]{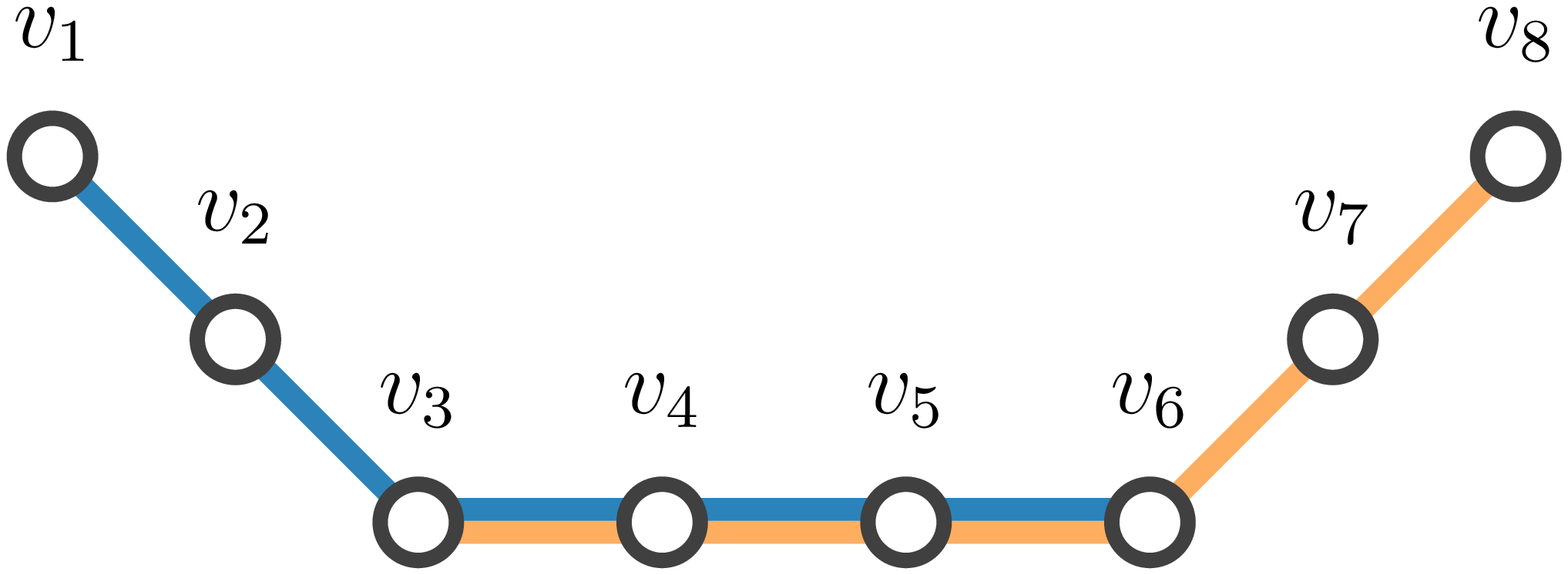}
    \caption{Path-based support of $\Pi_2$.}
  \end{subfigure}
  \par\bigskip
  \begin{subfigure}[b]{0.8\columnwidth}
  \small
  \centering
\begin{tabular}{c|*{8}{c}}
     &$v_3$ & $v_1$ & $v_7$ & $v_4$ &$v_5$ & $v_2$ & $v_8$ & $v_6$ \\ \hline
    $s_1$ & 1    & 1    & 0    & 1   & 1    & 1    & 0   & 1      \\ %
    $s_2$ & 1   & 0    & 1    & 1   & 1    & 0    & 1    & 1     \\ %
    \end{tabular}
    \caption{Incidence matrix of $\Pi_1$.}
  \end{subfigure}
  \par\bigskip
  \begin{subfigure}[b]{0.8\columnwidth}
  \small
  \centering
\begin{tabular}{c|*{8}{c}}
 &$v_1$ & $v_2$ & $v_3$ & $v_4$ &$v_5$ & $v_6$ & $v_7$ & $v_8$ \\ \hline
$s_1$ & 1    & 1    & 1    & 1   & 1    & 1    & 0   & 0      \\ %
$s_2$ & 0    & 0    & 1    & 1   & 1    & 1    & 1    & 1     \\ %
\end{tabular}
\caption{Incidence matrix of $\Pi_2$.}
  \end{subfigure}
  \caption{Consecutive ones property. Both (a) and (b) are valid support graphs. (a) shows the path-based support of incidence matrix (c), while (b) shows the support graph 
  of incidence matrix (d). Permutation $\Pi_2$ with  incidence matrix (d) has the consecutive ones property.} 
  \label{fig:c1pschema} 
\end{figure}

The second algorithm extracts a support graph by finding a permutation of the vertices, %
which maximizes the conjointness of metro lines. 
Two metro lines are {\em conjoint} if they are drawn as a pair of parallel lines between two vertices, thus reducing the number of edges in the support graph. 
While it is NP-complete to find a support graph with a minimal number of edges~\cite{bcps-psh-12}, 
the number of edges in the graph inversely correlates with the consecutive ones property of the incidence matrix $A=(a_{ij})$, %
where an entry $a_{ij} = 1$ iff the vertex $v_i$ is in the hyperedge $E_j$. 
A matrix has the {\em consecutive ones propert}y if there exists a permutation of its columns such that all non-zero elements of each row appear consecutively; see  \autoref{fig:c1pschema}. 

This is important because if an incidence matrix has the consecutive ones property, then there is a planar support graph~\cite{ks-cmoct-03}. Therefore, we try to minimize the violation of the consecutive ones property. 
To solve the problem of finding an optimal or near-optimal permutation of the vertices we model the consecutive ones property as a TSP instance, where the resulting optimal route is the permutation of vertices of the support graph. For the calculation of the cost matrix $C=(c_{ij})$, we consider each column $i$ in the incidence matrix as a vector $a_i$. The cost of traversing from vertex $v_i$ to $v_j$ in the TSP is $c_{ij} = \left\lVert  a_i - a_j \right\rVert$. 
We additionally add a dummy vertex whose column vector is $\overrightarrow{0}$, which functions as the initial or terminal vertex of the route. This is necessary as otherwise we would incorrectly ignore the cost of the last vertex in the route. 

The implementation for finding an exact solution uses the mixed-integer-programming (MIP) problem formulation of Miller et al.~\cite{miller1960integer}. We constrain the maximal time for finding a solution by stopping the MIP solver after too much time has passed and falling back to using the simulated annealing heuristic to find an approximate solution.

For each hyperedge, we can then construct a path by beginning at the dummy vertex we introduced and traversing the tour, adding vertices as we encounter them. The union of all of these paths then becomes the support graph.

\subsection{Expansion of Condensed Graph (Step 2)}

Once the support graph has been created, we re-introduce the vertices that were removed during the earlier simplification. This involves two steps: expanding the vertices that were condensed into a single vertex (see \autoref{fig:expand}) and inserting the vertices which only belonged to a single-set. The first step is straight-forward, but we propose two different methods for inserting single-set vertices.

\subsubsection{First Viable}

With this method, all single-set vertices are simply prepended to the path corresponding to their hyperedge; see \autoref{fig:firstViable}. This approach can be aesthetically appealing because it mimics the structure of many real-world metro systems, with a dense cluster of interconnected stations near the middle of the graph in the city center, and long, solitary lines leading to the suburbs.
Problems can arise with larger datasets, however. If the cluster of stations in the center of the metro map is too large, or its vertices are of too high degree, then it becomes extremely difficult to draw the map in an octolinear style, and the resulting layout is often both difficult to read and only vaguely reminiscent of a metro map. This is the motivation behind our second insertion algorithm.

\subsubsection{Split Insert}

\begin{figure}[b!]
  \centering
  \begin{subfigure}[b]{\columnwidth}
    \centering
    \includegraphics[width=\linewidth]{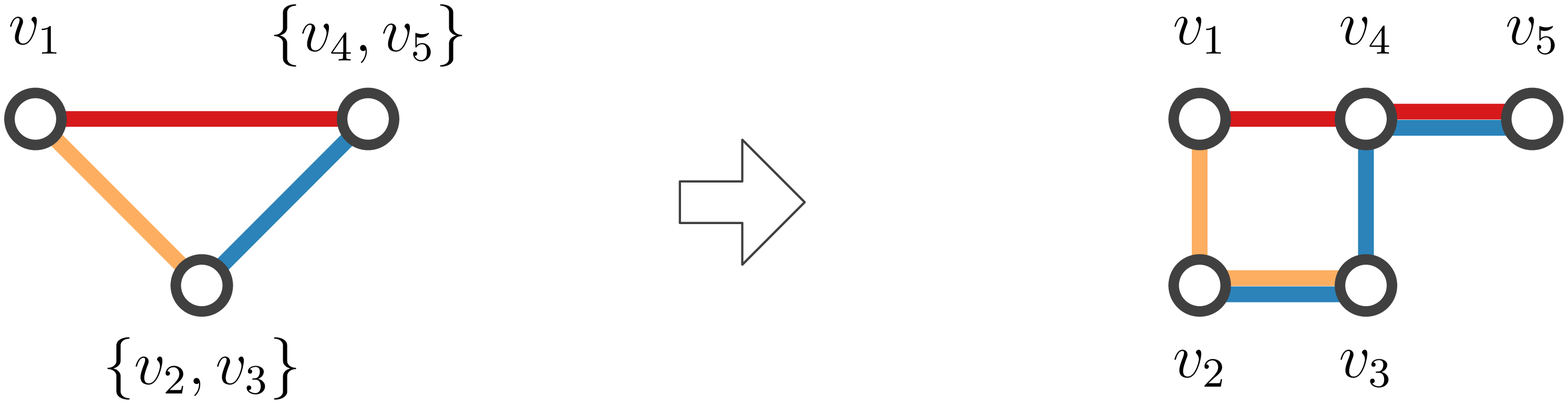}
    \caption{Expansion of merged vertices}
    \label{fig:expand}
  \end{subfigure}
  \centering
  \begin{subfigure}[b]{0.49\columnwidth}
    \centering
    \includegraphics[width=\linewidth]{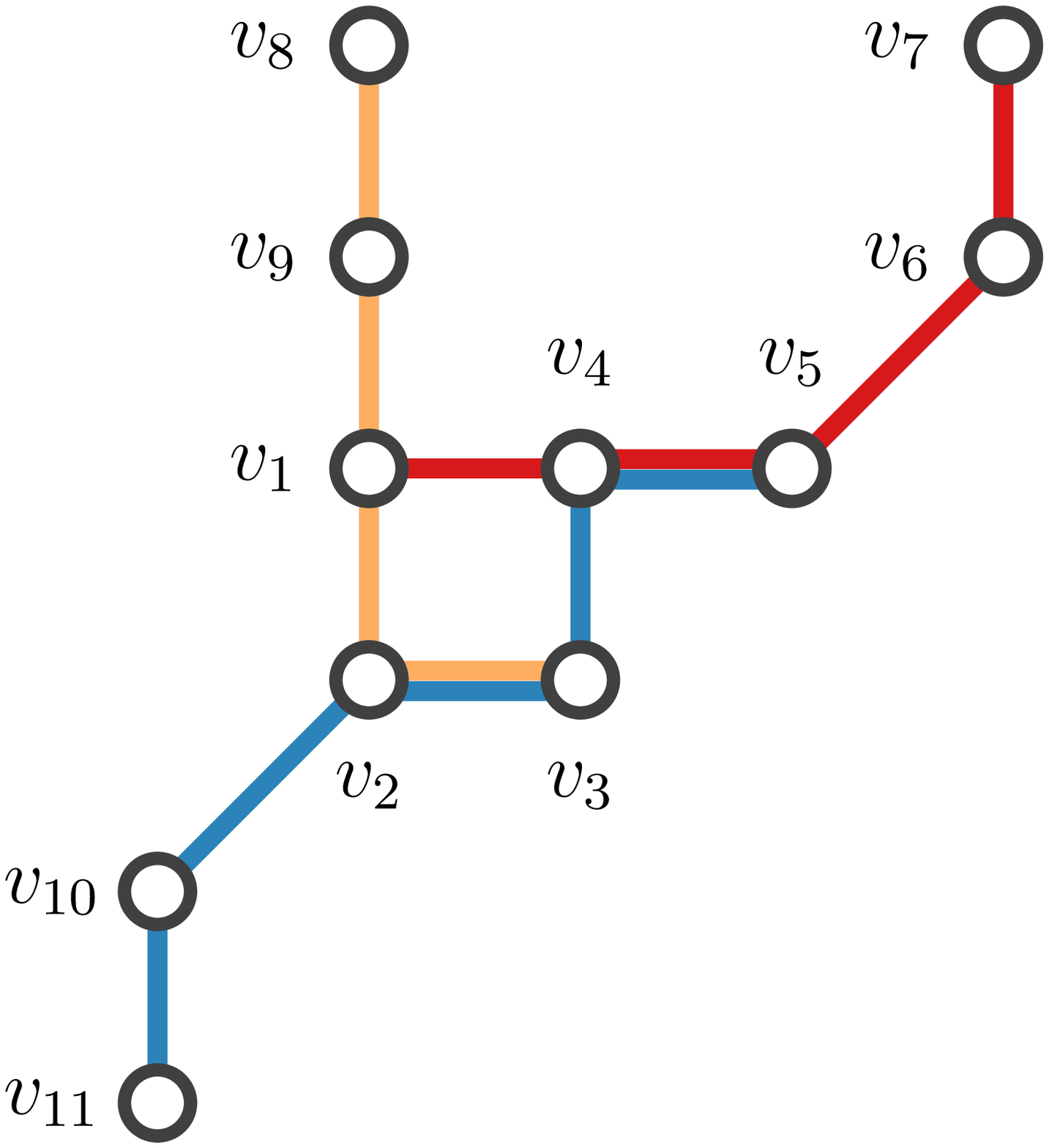}
    \caption{First Viable}
    \label{fig:firstViable}
  \end{subfigure}
  \begin{subfigure}[b]{0.49\columnwidth}
    \centering
    \includegraphics[width=\linewidth]{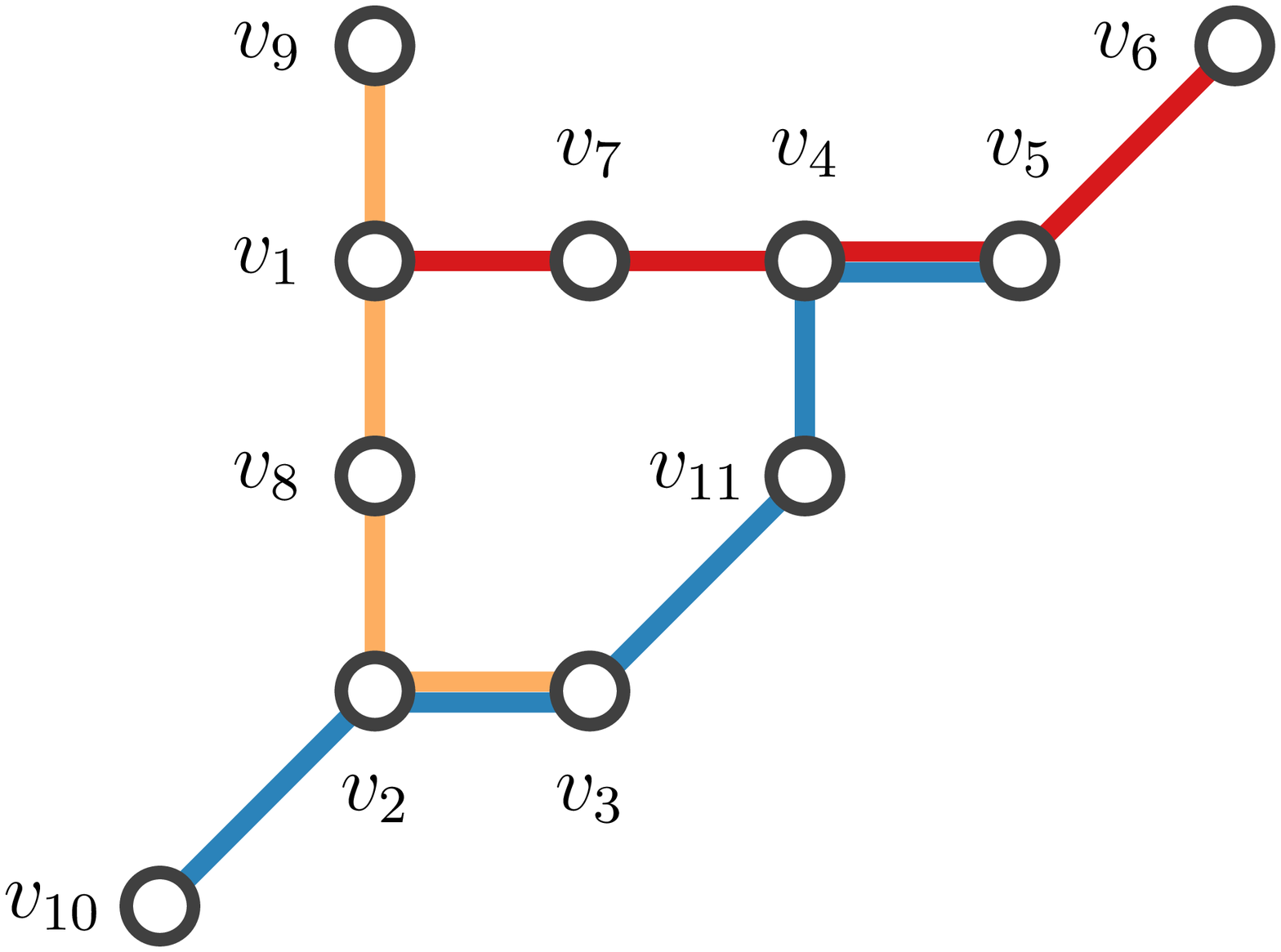}
    \caption{Split Insert}
    \label{fig:splitInsert}
  \end{subfigure}
  \caption{\new{Support Graph before and after vertex expansion (a). The same support graph after the single-set vertex insertion (b) and (c).}}
\end{figure}

With this method, we also insert single-set vertices inside our paths, and not only at the periphery. The possible candidates for locations to insert are all of the edges which are traversed only by the given hyperedge. Inserting single set vertices into the path at these locations has the effect of spreading the center of the graph and spacing high-degree vertices further from each other. This leads to a more readable layout, and one which more closely resembles a metro map; see \autoref{fig:splitInsert}.
In order to preserve some of the aesthetic advantages of the first viable insertion method, we conserve half of the single-set vertices to place at the beginning of the path, and distribute the rest evenly along all of the candidate edges.

\new{We tried other strategies for this step, including maximizing the girth of the resulting graph, and placing vertices at both ends of the path. The results were somewhat mixed: more space efficient in some cases, or introducing edge crossings and decreasing readability in other cases. 
}

\subsection{Initial Layout and Refinement (Step 3)}

The input to this step is a path-based support graph $G = (V,E)$. Our goal is to create an initial embedding of this graph into the plane, 
and we propose two methods for constructing this embedding.

\subsubsection{Repeated Refinement of Paths}

As the name suggests, this method refines paths by repeatedly reordering the vertices along them. \new{We implement two different pipeline options which follow the same essential logic, but utilize different pre-existing layout algorithms as a basis. One uses the Kamada-Kawai algorithm~\cite{kamada1989algorithm} as implemented in NetworkX~\cite{hagberg2008exploring} and the other uses Neato from graphviz~\cite{ellson2001graphviz}}.
We begin by creating an initial layout of the support graph. Each hyperedge in the initial input corresponds to a path in $G$. However, these paths can initially meander and self-intersect and so we reorder the vertices along them.

Once again, we treat this as a TSP problem. This time, however, we are not merely concerned with the similarity score between two vertices; we also want to take into account the Euclidean distance between vertices in our initial embedding. The numeric ranges of these two values can vary wildly, so we say that the cost of an edge between two vertices is the geometric mean of the two.

After we solve the TSP problem for each of the hyperedges, we once again take the union of those paths as the new path-based support graph and compute a new layout. 
The  result is a new embedded graph whose paths are better behaved. This process can be repeated as many times as needed until the paths stabilize.

It is highly desirable to avoid  self-intersecting paths, as this can be confusing and it does not normally occur in metro maps.
One of the advantages of the two-opt heuristic is that it naturally eliminates most self-crossings, provided the underlying space is metric \cite{johnson1997traveling}. 
To take advantage of this property, on the final iteration, we drop the similarity score and base costs purely on Euclidean distance.

\subsubsection{Force-Based Initial Layout}

This method relies on a modified implementation of the Fruchterman-Reingold spring embedder algorithm~\cite{fruchterman1991graph}.
To improve performance and reach faster convergence to an equilibrium state, we implemented an adaptive temperature scheme~\cite{frick1994fast}. %
The ideal edge length is set to a constant target length and the drawing area is unrestricted. We set an iteration threshold of 5000, after which we terminate the algorithm even if it has not converged to an equilibrium state before.
We provide two options for initial positions of the vertices: random positions and positions computed via multidimensional scaling (MDS) using the graph theoretic distances~\cite{cox2008multidimensional}. MDS requires fewer iterations for convergence and is the default option. %

\subsection{Schematization (Step 4)}

At this stage we have an embedded path-based support graph and we modify its layout to obtain uniform edge lengths and to
impose the constraint of \textit{octolinearity}: each edge should have a slope which is a multiple of $45^\circ$. We offer two methods for schematization.

\subsubsection{Least Squares Approximation}

The first schematization method is adapted from the work of Lutz et al.~\cite{lutz2014realtime}. 
The goal of schematization is to produce a layout, where every edge has a uniform length and is drawn octolinearly. This is not always possible (e.g., a triangle cannot be realized). 
We treat edge uniformity and octolinearity as soft constraints and minimize the error. However, the error function is non-linear, making it difficult and time-consuming to optimize. To overcome this, we create a linear approximation of the problem which can then be solved quickly. %

To create an approximation, we first choose a desired angle for each edge. This is done by creating eight `ports' around each vertex, separated by $45^\circ$ angles. We treat the assignment of edges (incident to the vertex) to ports as a discrete least squares problem, with the goal of minimizing the discrepancy between port assignments and original edge directions. This gives us the angle at which we would like each edge to leave the vertex.
Next, for each edge, we have two cost functions representing error in edge length and error in angle. As these functions are nonlinear, we approximate them using scalar projection. These new, linear constraints are then optimized using least squares approximation, giving us an optimal position for each vertex.

\subsubsection{Force-Based Schematization}

The second schematization approach %
follows the force-based metro map layout algorithm by Chivers and Rodgers~\cite{cr-oflwmpsd-14}. The idea is to use a standard spring embedder and add magnetic forces that pull edges towards an octolinear position. 
While the spring embedder forces, $f_\text{spring}$ are identical to the ones discussed earlier, the magnetic forces $f_\text{mag}$ are new and are also calculated for each edge of the graph at every iteration. 
For every edge the closest angle to one of the eight possible octolinear directions is calculated. Then the ideal octolinear position of one of the incident vertices is determined by rotating the edge according to the previously calculated angle around its midpoint. 
The vector between this vertex and its ideal octolinear position is used to calculate the magnetic force. 
To keep the layout from collapsing, $f_\text{mag}$ also adds a repulsive or attractive force to both incident vertices that either pushes the vertices apart or pulls them together, depending on the actual length of the edge compared to the ideal edge length.

We also define two weights $\alpha_\text{spring}$ and $\alpha_\text{mag}$, which are used as scaling factors for calculating the resulting sum of forces before the displacement is determined. Adaptive temperature~\cite{frick1994fast} is used to counter oscillating movements.

The Chivers-Rodgers approach has multiple stages and in each stage $\alpha_\text{spring}$ and $\alpha_\text{mag}$ vary. %
The first stage is identical to a spring embedder
and since the input to Step 4 of our pipeline already provides an initial layout, we skip this stage.
In the second stage we perform a fixed number of iterations, while linearly decreasing $\alpha_\text{spring}$ from $1.0$ to $0.0$ and linearly increasing $\alpha_\text{mag}$ from $0.0$ to $1.0$, such that $\alpha_\text{mag} + \alpha_\text{spring} = 1.0$. 
In the third stage we disable the spring forces and only apply magnetic forces to each vertex.
We set the maximum number of iterations to 700 for stage 2 and to 200 for stage 3. In each iteration we calculate the remaining energy in the system and terminate if it falls below a convergence threshold.

\subsection{Postprocessing}

To further enhance the visualization of the hypergraph, we apply two postprocessing steps to the schematized layout. First, we minimize {\em line crossings} by considering the order of incoming and outgoing lines for each vertex, such that crossovers are minimized, which leads to less visual complexity and improves the continuity of lines. Second, we apply a station labeling algorithm, to generate metro-like labels with the name of each vertex.

\subsubsection{Line Crossing Minimization}

\begin{figure}[t]
  \centering
  \begin{subfigure}[b]{0.49\columnwidth}
    \centering
    \includegraphics[width=\linewidth]{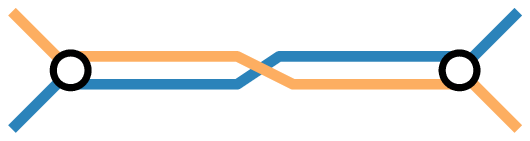}
    \caption{Unavoidable line crossing}
  \end{subfigure}
  \begin{subfigure}[b]{0.49\columnwidth}
    \centering
    \includegraphics[width=\linewidth]{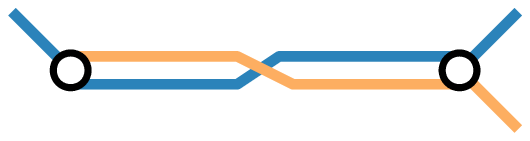}
    \caption{Avoidable line crossing}
  \end{subfigure}
  \caption{It is often impossible to create a drawing with no line crossings. This figure illustrates some of the different situations that can arise.}
  \label{fig:linecrossing}
\end{figure}

Finding orderings that minimize the number of crossings is known as the \textit{metro-line crossing minimization} (MLCM) problem; see \autoref{fig:linecrossing}. We add the additional requirement that terminating lines must be at the leftmost or rightmost side of the edge leading up to their terminus, thus preventing gaps between continuing lines. 
Under this requirement, referred to as the periphery condition, %
the MLCM problem is NP-hard~\cite{bekos2007line}. However, if we are given the sides on which each line should terminate as part of our input, called \textit{terminator positions}, then the problem can be solved efficiently in polynomial time~\cite{nollenburg2010linecrossings}.

We therefore employ a heuristic algorithm developed by Asquith et al.~\cite{asquith2008metroilp}, which assigns terminator positions locally. 
The central idea of this algorithm is as follows: for each terminator position, we ask how many crossings will definitely occur if we choose to place the terminator on the left ($f_L$), how many crossings will definitely occur if we place it on the right ($f_R$), and how many crossings might occur either way based on other terminators whose positions have not yet been determined ($r$).
Once we have calculated these three numbers for every terminator, we then iteratively fix the position of the next terminator where we are most certain we can make an optimal choice, i.e., the terminator where $\left| f_L - f_R \right| - r$ is maximized. 
Once we have fixed the position of every terminator, we can employ the algorithm of N\"ollenburg~\cite{nollenburg2010linecrossings} to calculate line orders for every edge without introducing any new crossings.

\subsubsection{Labeling}

Before we render the output, we label all vertices. 
A label in our use case is a short piece of text. 
By following standard cartographic label placement principles~\cite{i-pnm-75,niedermann2018algorithmic}, we want to place each label close to its assigned vertex, so that it is clear which label belongs to which vertex. 
The labeling should avoid line-label and label-label overlaps to decrease visual complexity and avoid ambiguity. Each label must be either horizontal or have an angle of $\pm45^\circ$. Labels along a line segment should be placed on the same side and should have the same direction. To accomplish these goals, we implement the greedy labeling algorithm of Niedermann and Haunert~\cite{niedermann2018algorithmic}. The basic idea is this: first, each vertex gets candidate labels assigned. \new{Then, the algorithm scales all candidates and tries to find the largest font size, between 8 and 60 pts., such that there are no line-label or label-label intersections (called a \emph{valid} labeling)}. Finally, the algorithm greedily picks the best valid labels from all candidates.
Best in the sense that candidate labels have predefined weights assigned, which in turn are used as optimization target. We define lower weights for labels that are horizontal and expand to the right of the vertex if a line is vertical or diagonal. For horizontal lines we prefer labels with an angle of $\pm45^\circ$ which expand to the right of the vertex. \new{We abbreviate labels that are longer than 16 characters during scaling in order to help find a valid labeling.  Unabbreviated labels are shown when their placement permits a valid labeling.}

\new{If no valid labeling can be found for the smallest scale allowed, our system defaults to a fallback labeling, where we use the following simple scheme to label each vertex. Vertices with horizontal edge direction are assigned labels at $45^\circ$ angle. In all other cases we place the label below the vertex in horizontal direction. In the fallback labeling, line-label and label-label intersections may occur.}

\subsection{Preset Pipelines}

\begin{figure}[t]
  \centering
  \begin{subfigure}[b]{0.51\columnwidth}
    \centering
    \includegraphics[width=\linewidth]{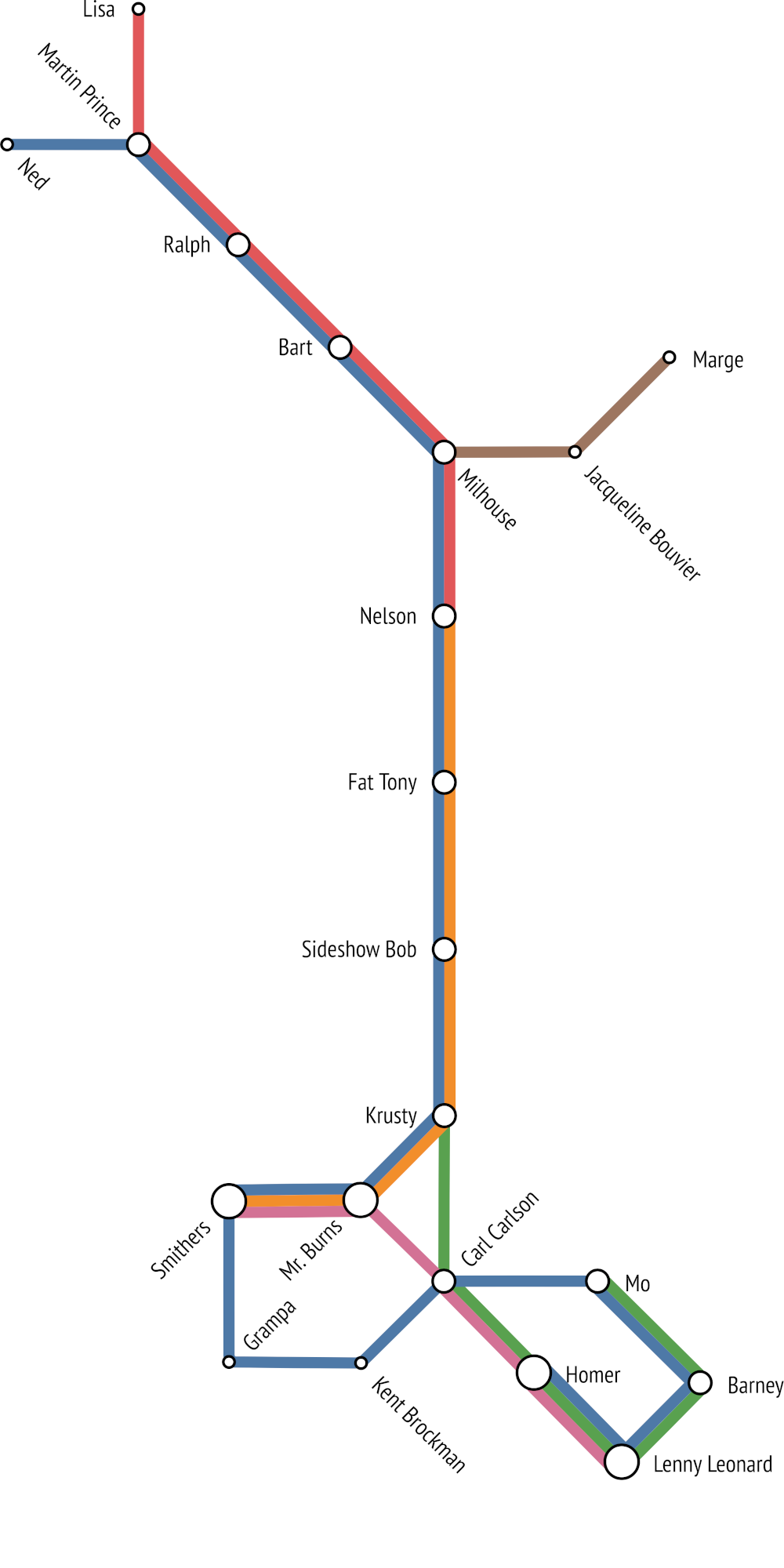}
    \caption{Preset: Balanced}
    \label{fig:presetsSimpsons1}
  \end{subfigure}
  \hfill
  \begin{subfigure}[b]{0.47\columnwidth}
    \centering
    \includegraphics[width=\linewidth]{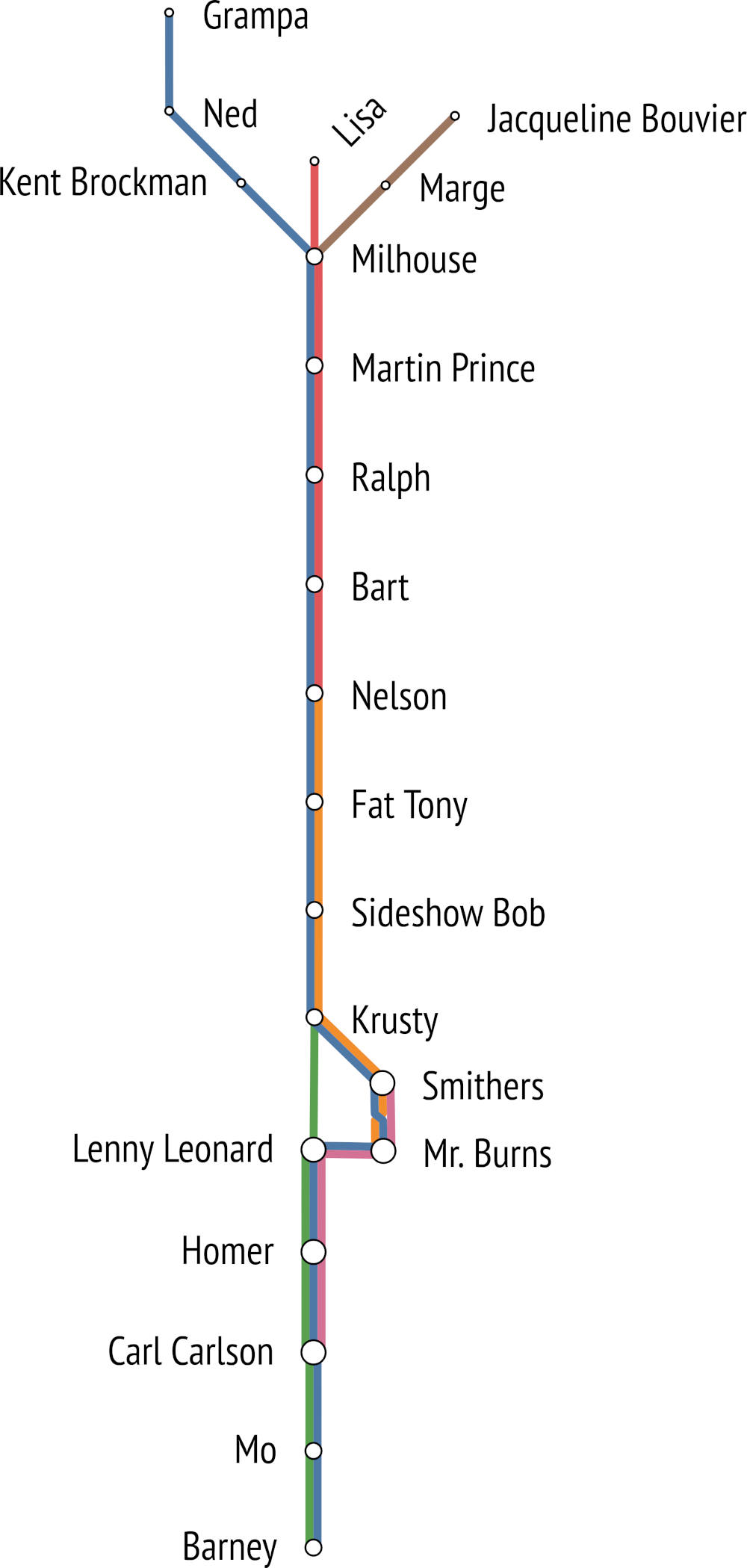}
    \caption{Preset: Simplicity}
    \label{fig:presetsSimpsons2}
  \end{subfigure}
  \begin{subfigure}[b]{0.9\columnwidth}
    \centering
    \includegraphics[width=\linewidth]{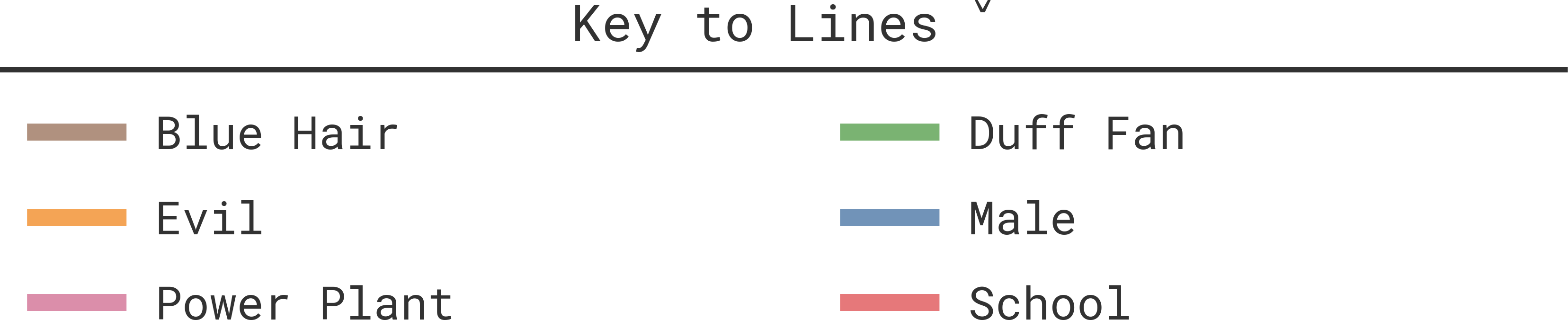}
  \end{subfigure}
  \caption{\new{The Simpsons data set visualized with two presets.}}
\end{figure}

We performed extensive experimentation evaluating the different possible combinations of the pipeline steps described here. On the basis of our results, described in Section~\ref{sec:eval}, we provide the following three pipeline presets on the MetroSets website.

The \textbf{Balanced} preset is the default to balance speed, metro-map-likeness, and scalability. We believe it should be the first choice of pipeline for most tasks. It uses the TSP support graph algorithm, Split Insert insertion algorithm, TSP (Kamada-Kawai) layout algorithm, and Chivers-Rodgers schematization algorithm; see \autoref{fig:presetsSimpsons1}.

The \textbf{Max Speed} preset optimizes running time, and makes use of the most efficient pipeline option at every step. It is recommended for larger datasets or when fast running time is a priority. It uses the TSP support graph algorithm, Split Insert insertion algorithm, TSP (Neato) layout algorithm, and Least Squares schematization algorithm.

The \textbf{Simplicity} preset prioritizes visual simplicity of the map and produces maps with long, simple `suburb' lines and lines which run in parallel. 
It makes use of the Consecutive Ones support graph algorithm, First Viable insertion algorithm, Spring Embedder layout algorithm, and Chivers-Rodgers schematization algorithm; see \autoref{fig:presetsSimpsons2}.

\section{Map Rendering and Interface}\label{sec:design}

\subsection{Metro Map Rendering}

\new{Each vertex is drawn as a circle with diameter determined by either the number of incident sets or the largest number of adjacent lines of all incident edges. By default we use the latter option, but this can be changed while viewing a map.}

We use straight lines to connect vertices, with no space between parallel lines. Line crossing between two incident vertices are drawn in the center of the respective edge. The width of a line is fixed at 8 pts. \new{We use the Tableau20 color scheme as a categorical scheme to  color the metro lines. Thus for hypergraphs with more than 20 hyperedges, colors must be reused. We try to assign perceptually different colors to lines with the help of a greedy heuristic to optimize color difference between adjacent lines. Note that the underlying problem, known as max differential coloring, is NP-hard~\cite{bekos2017maximum}.}

The legend depicts which hyperedge is assigned which color and is placed as a collapsing overlay in the bottom right corner of the map.
The background of the drawing is light grey, to lessen eye strain.%

Before rendering the schematic map we scale the layout, such that the average edge length is 50 pts. The size of the output map also determines the scaling range for the labels.

\subsection{MetroSets Interface}

In addition to the static rendering of the schematization, we provide several means to explore the data via interactions. \new{According to the task taxonomy of Alsallakh et al.~\cite{alsallakh2014visualizing}, visualization of set-typed data is concerned with performing tasks grouped into three categories}. Our visualization design mainly targets two of those categories, namely tasks related to elements and tasks related to sets and set relations. Of the 21 tasks over those two categories, our system supports 17, of which 11 are possible without interaction. Eight tasks can be done visually, but are enhanced with interactivity. The exact list can be found in \autoref{tab:interactivity} in the supplementary materials. We did not consider the third task category, tasks related to element attributes.

Following Shneiderman's mantra \cite{shneiderman1996eyes}, ``overview first, zoom and filter, then details on demand,'' we scale the visualized schematic drawing to the maximal available screen space while keeping the aspect ratio fixed. Zoom and pan  can be used to focus on different parts of the visualization. For details-on-demand we use hover interaction to emphasize specific sets or elements, while the set operation mode can be used to perform standard set operations and filter the data accordingly. 

In the hover mode, lines and vertices can be hovered to emphasize them. When emphasizing a line or vertex, we apply low opacity to all other lines and vertices which are not in focus. This makes it possible to focus on one part of the visualization while keeping the overall structure of the map visible. When hovering a line, we set the focus on the line itself and all vertices that are incident to the line. When hovering a vertex, we set the focus on all hyperedges the vertex is incident to, as well as all other vertices this set of hyperedges covers. We also show a tool-tip with information such as vertex label and additional data attributes. All interactions that apply to lines can also be triggered by hovering or clicking the entries of the legend. All labels that are abbreviated (because their length is above the 16 character limit) can be hovered to show the full text.

The data can be further filtered with the following set operations:

\begin{itemize}[noitemsep]
    \item The intersection mode can be used to show set intersections. Hyperedges can be selected and deselected to focus on the set of vertices that are incident to all of the selected hyperedges. 
    \item The union mode can be used to focus the selected hyperedges along with all vertices incident to at least one of them.
    \item The complement mode emphasizes all of the vertices which do not belong the union of the selected hyperedges.
    \item The \new{symmetric difference} mode emphasizes vertices that are covered by the union of the selected hyperedges while being complementary to the intersection of all selected hyperedges.
    \item Finally, the subtract mode can emphasize the difference between sets. The operation works on two levels. First: by clicking on lines a union of all selected hyperedges is constructed. Second: when additionally a  modifier key is pressed when a line is clicked, an independent second union of hyperedges is created. After each interaction the second union is subtracted from the first union and the remaining vertices are highlighted.  
\end{itemize}

\begin{figure}[t]
    \centering
    \includegraphics[width=\columnwidth]{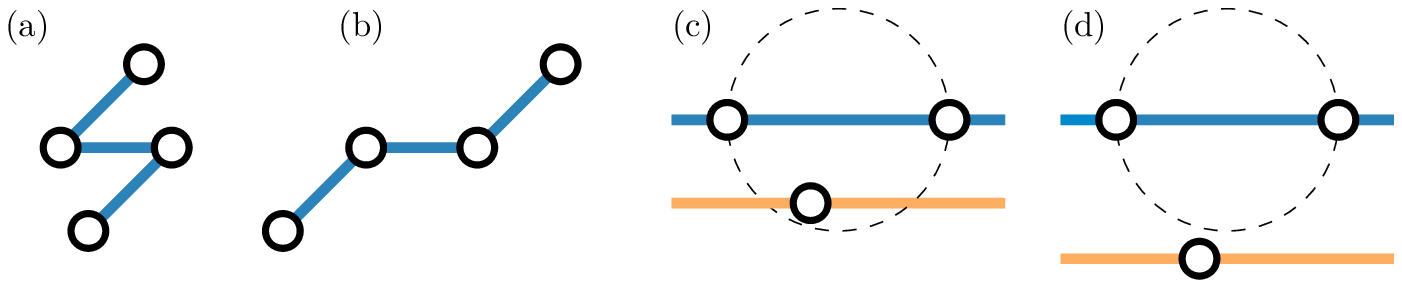}
    \caption{Monotonicity is a measure of a line's tendency to reverse direction: (a) has a score of 2, while (b) has a score of 0. Gabriel Score is a measure of how well-spaced unrelated elements are: (c) has a score of 1, while (d) has a score of 0.}
    \label{fig:qualityMeasurements}
\end{figure}

\section{Evaluation}\label{sec:eval}

\subsection{Quality Measures}\label{sub:qualitymeasures}

To evaluate the maps produced by the various combinations of pipeline steps, we have created a library of methods to calculate quality metrics that correspond to the design goals laid out in section \ref{sec:design}, which
correspond to generally well accepted principles of schematic metro map design \cite{nollenburg2014survey} \cite{roberts2017preference}. These should be taken as our best preliminary attempt to capture the concept of ``a good metro-map'' in a rigorous and quantifiable way which is amenable to large-scale testing. The following are the primary statistics we calculate. In the supplementary materials, we provide examples of maps which are similar in most respects but differ strongly in one metric as a way of illustrating the metric's impact; see \autoref{fig:metriccomparison}.

\textit{Octolinearity} is one of the most immediately recognizable features of a metro-map. A highly octolinear schematization maximizes the coherence of the visualization through the presence of many parallel lines~\cite{roberts2017preference}, and is identified by N\"ollenburg as one of the primary rules of metro-map generation \cite{nollenburg2014survey}. We quantify the octolinearity of a map through the average and maximum number of degrees that each edge differs from a multiple of 45 degrees.

\textit{Edge Uniformity} is the principle that each edge should have a uniform length. High edge uniformity helps to enlarge and focus densely interconnected regions of the graph, enhancing readability, and is identified explicitly as a rule of metro-map generation by N\"ollenburg~\cite{nollenburg2014survey}. We quantify both the average and the maximum proportion of each edge to the average edge length.

\textit{Monotonicity} of a line is a measure of its tendency to change direction. Optimizing monotonicity simplifies the visualization and appeals to the gestalt principle of continuity, reducing the difficulty of visually tracking lines. We quantify monotonicity on a per-line basis by calculating an ``As the crow flies" vector from the first vertex on the line to the last, and then counting the number of times the edges of the line reverse orientation with respect to this direct vector using dot products. The monotonicity of the graph is the sum of the monotonicity of its lines, and a score of 0 is optimal; see Figure~\ref{fig:qualityMeasurements}.

\textit{Gabriel Score} is  inspired by the Gabriel graph~\cite{gabriel1969new}, in which two vertices $p$ and $q$ are connected by an edge if and only if the circle with diameter $\overline{pq}$ contains no other vertices. The Gabriel Score of a metro map  is computed by counting the number of times this condition fails to hold, i.e., the number of (vertex, edge) pairs such that the vertex is not incident to the edge, but lies within the circle defined by the edge. 
The Gabriel score measures the ``station separation" design goal, motivated by the notion that unrelated features should be kept far apart; see Figure~\ref{fig:qualityMeasurements}.

\textit{Consecutive Ones} is a property that measures the ``conjointness/closeness" design goal, motivated by the idea that, wherever possible, the intersection of two hyperedges should form a contiguous path. In this way, the fact that the two hyperedges overlap will be manifested in a visually conspicuous manner as two lines running in parallel. We quantify this property by finding the subgraph induced by each pair of metro lines and counting the number of connected components; each component beyond the first increases the score of the graph by 1, with a score of 0 being optimal.

{\em Edge crossings, self crossings, and line crossings} observed in the final graph measure the ``planarity" and ``edge/line-crossing minimization" design goals. These scores are just counts of the different types of intersections.

{\em Running time} is the final measure, as 
this is important for the usability and responsiveness of the online MetroSets system.

\subsection{Datasets}

To test our system, we created a tool for randomly extracting sub-hypergraphs. \new{Given a specified number of nodes and hyperedges, our tool first selects a random sub-hypergraph with the given number of hyperedges, and then greedily replaces the hyperedges with others from the dataset until it also has the correct number of nodes.} For our base dataset, we used a large (6,714 vertices, 39,774 hyperedges) hypergraph of recipes and ingredients drawn from the Kaggle ``What's Cooking?" competition \cite{Amburg-2020-categorical}. Then, for each value $n \in 20,25,...160$ and $h \in 6,7,...,19$, we extracted ten sub-hypergraphs with $n$ vertices and $h$ hyperedges (for a total of $29\times 14\times 10=4060$ hypergraphs).

\subsection{Experimental Design}
To quantify how the choice of pipeline steps affects our quality metrics, we chose a restricted subset of our datasets ($n \in 40,80,...,160$, $h \in 6,9,...,18$, with 3 datasets per combination of $n$ and $h$) and fed each of them through every combination of pipeline steps. Then, for each map produced, we recorded descriptive statistics indicating its performance by each of the quality measures described in Section~\ref{sub:qualitymeasures}.

Treating the number of vertices and number of hyperedges as categorical variables, we performed a 6-way ANOVA test \cite{fisher1992statistical} for each of our descriptive statistics, including interaction terms for all pairs of independent variables. For each statistically significant factor, we performed additional post-hoc analysis using Tukey's test\cite{tukey1949comparing}, and quantified its impact with the Omega Squared measure for effect size\cite{hays1963statisticsfor}.

\begin{table}[tb]
\setlength{\tabcolsep}{3pt}
 \small
 \begin{tabularx}{\columnwidth}{c X|*{11}{X}}
 \multicolumn{1}{c}{Factor\rule{0pt}{10ex}} && \begin{rotate}{60}  Gabriel  Score\end{rotate} & \begin{rotate}{60}  Consecutive  Ones\end{rotate} & \begin{rotate}{60}  Monotonicity\end{rotate} & \begin{rotate}{60}  Avg.  Octilinearity\end{rotate} & \begin{rotate}{60}  Max  Octilinearity\end{rotate} & \begin{rotate}{60}  Edge  Crossings\end{rotate} & \begin{rotate}{60}  Self  Crossings\end{rotate} & \begin{rotate}{60}  Line  Crossings\end{rotate} & \begin{rotate}{60}  Running  Time\end{rotate} & \begin{rotate}{60}  Avg.  Edge  Uni.\end{rotate} & \begin{rotate}{60}  Max  Edge  Uni.\end{rotate} \\
 \hline 
 Insert&  p    \newline  $\omega^2$ &   0.00  \newline  0.01 &   0.47  \newline  0.00 & \pg  0.00  \newline  0.02 &   0.22  \newline  0.00 &   0.00  \newline  0.01 &   0.90  \newline  0.00 &   0.00  \newline  0.00 &   0.18  \newline  0.00 &   0.00  \newline  0.00 &   0.00  \newline  0.00 &   0.00  \newline  0.00 \\
 \hline 
 Layout&  p    \newline  $\omega^2$ & \pg  0.00  \newline  0.05 &   0.85  \newline  0.00 & \pg  0.00  \newline  0.04 & \pg  0.00  \newline  0.02 &   0.00  \newline  0.00 & \pg  0.00  \newline  0.04 & \dg  0.00  \newline  0.20 &   0.04  \newline  0.00 & \dg  0.00  \newline  0.19 & \pg  0.00  \newline  0.04 & \pg  0.00  \newline  0.03 \\
 \hline 
 $|\hE|$&  p    \newline  $\omega^2$ & \dg  0.00  \newline  0.37 & \dg  0.00  \newline  0.18 & \dg  0.00  \newline  0.42 & \dg  0.00  \newline  0.22 & \dg  0.00  \newline  0.29 & \dg  0.00  \newline  0.40 & \mg  0.00  \newline  0.08 & \dg  0.00  \newline  0.37 &   0.00  \newline  0.00 & \dg  0.00  \newline  0.47 & \dg  0.00  \newline  0.19 \\
 \hline 
 $|\hV|$&  p    \newline  $\omega^2$ & \dg  0.00  \newline  0.16 & \mg  0.00  \newline  0.06 & \mg  0.00  \newline  0.08 & \pg  0.00  \newline  0.02 & \pg  0.00  \newline  0.03 & \mg  0.00  \newline  0.12 &   0.00  \newline  0.01 & \pg  0.00  \newline  0.01 & \dg  0.00  \newline  0.59 & \mg  0.00  \newline  0.08 & \dg  0.00  \newline  0.23 \\
 \hline 
 Schem.&  p    \newline  $\omega^2$ &   0.58  \newline  0.00 &   0.94  \newline  0.00 &   0.00  \newline  0.00 &   0.00  \newline  0.01 & \dg  0.00  \newline  0.17 &   0.04  \newline  0.00 &   0.84  \newline  0.00 &   0.87  \newline  0.00 & \pg  0.00  \newline  0.01 &   0.00  \newline  0.00 & \mg  0.00  \newline  0.12 \\
 \hline 
 Support&  p    \newline  $\omega^2$ & \pg  0.00  \newline  0.02 &   0.06  \newline  0.00 &   0.00  \newline  0.00 &   0.00  \newline  0.00 &   0.32  \newline  0.00 &   0.00  \newline  0.01 &   0.10  \newline  0.00 &   0.00  \newline  0.01 &   0.00  \newline  0.00 &   0.00  \newline  0.00 &   0.95  \newline  0.00 \\
 \hline 
 Layout,$|\hE|$&  p    \newline  $\omega^2$ & \pg  0.00  \newline  0.03 &   0.07  \newline  0.00 & \pg  0.00  \newline  0.02 &   0.00  \newline  0.01 &   0.16  \newline  0.00 & \pg  0.00  \newline  0.04 & \mg  0.00  \newline  0.12 &   0.01  \newline  0.00 &   0.00  \newline  0.00 & \pg  0.00  \newline  0.01 & \pg  0.00  \newline  0.01 \\
 \hline 
 Layout,$|\hV|$&  p    \newline  $\omega^2$ & \pg  0.00  \newline  0.01 &   0.01  \newline  0.00 &   0.38  \newline  0.00 &   0.00  \newline  0.00 &   0.02  \newline  0.00 &   0.00  \newline  0.00 & \pg  0.00  \newline  0.01 &   0.36  \newline  0.00 & \dg  0.00  \newline  0.16 &   0.13  \newline  0.00 &   0.03  \newline  0.00 \\
 \hline 
 $|\hE|$,$|\hV|$&  p    \newline  $\omega^2$ & \mg  0.00  \newline  0.11 & \dg  0.00  \newline  0.15 & \pg  0.00  \newline  0.04 & \pg  0.00  \newline  0.05 & \pg  0.00  \newline  0.02 & \mg  0.00  \newline  0.09 &   0.00  \newline  0.01 & \mg  0.00  \newline  0.09 &   0.00  \newline  0.00 & \pg  0.00  \newline  0.03 &   0.00  \newline  0.01 \\
 \hline 
 $|\hE|$,Schem.&  p    \newline  $\omega^2$ &   0.94  \newline  0.00 &   1.00  \newline  0.00 &   0.28  \newline  0.00 & \dg  0.00  \newline  0.17 & \pg  0.00  \newline  0.05 &   0.24  \newline  0.00 &   0.97  \newline  0.00 &   1.00  \newline  0.00 &   0.84  \newline  0.00 & \pg  0.00  \newline  0.06 & \pg  0.00  \newline  0.06 \\
 \hline 
 $|\hV|$,Schem.&  p    \newline  $\omega^2$ &   0.60  \newline  0.00 &   1.00  \newline  0.00 &   0.66  \newline  0.00 &   0.00  \newline  0.01 & \pg  0.00  \newline  0.02 &   0.66  \newline  0.00 &   0.93  \newline  0.00 &   1.00  \newline  0.00 &   0.00  \newline  0.01 & \mg  0.00  \newline  0.07 & \pg  0.00  \newline  0.02 \\
 \hline 
 \end{tabularx}
\caption{ \new{Abbreviated summary of the impact of different factors on each metric, based on the results of ANOVA tests. Each column represents a metric and each row represents a factor (or pair of factors). The top value in each cell is the p-value of the effect of the factor on the metric and the bottom value is the $\omega^2$ measure of effect size, with all values rounded to 2 decimal places. Additionally, a cell is colored} \textcolor{deepGreen}{deep}, \textcolor{medGreen}{medium}, or \textcolor{paleGreen}{light} \new{green if that factor has a large ($\omega^2 \geq 0.14$), medium ($0.06 \leq \omega^2 < 0.14$) or small ($0.01 \leq \omega^2<0.06$) effect on that metric, respectively. A cell is left white if that factor's effect is very small ($\omega^2 < 0.01$). Pairs of factors represent interaction terms; when a cell in a row corresponding to an interaction term is colored, it means that the value of that metric behaves differently from how you would guess based on looking at the two interacting factors in isolation. This table is a subset of the full results,
\iflabelexists{tab:anova-results2}
  {which are recorded in Table~\ref{tab:anova-results2}.}
  {available in the supplementary materials.} }}
\label{tab:anova-results}
\end{table}

\subsection{Results}

We briefly summarize our main findings:
\begin{itemize}[noitemsep]
    \item The running time of MetroSets is most strongly influenced by the number of vertices.
    \item For most metrics, the number of hyperedges is the most important factor. 
    \item For most metrics, the choice of support or insertion does not have a large effect on the final score, and the impact of the schematization algorithm is limited to octolinearity and edge uniformity.
    
\end{itemize}

We present the results of our ANOVA tests in \autoref{tab:anova-results}. The most important point to stress is that many of the interaction terms are significant. In other words, it is in general not possible to predict the quality of a map by looking at the input size or each pipeline step in isolation. For example, when optimizing the number of line crossings, the spring embedder is the worst performing layout algorithm when paired with the TSP support algorithm, but it performs best when paired with the C1P support algorithm. This is denoted in \autoref{tab:anova-results} by the (Layout,Support) cell being highlighted in the column for line crossings. Salient results from our post-hoc analysis include:

\begin{itemize}[noitemsep]
    \item Several metrics, such as Consecutive Ones and Average Octolinearity, improve with increased number of vertices. In other words, a larger dataset does not necessarily imply worse results.
    \item The two TSP layout algorithms perform better than the Spring Embedder for most tasks. The Spring Embedder can outperform them on some metrics when paired with Consecutive Ones.%
    \item The Consecutive Ones algorithm is slightly slower than the TSP support graph algorithm, Chivers-Rodgers is slightly slower than the Least Squares schematization algorithm, 
    and both TSP layout algorithms are faster than the Spring Embedder. 
\end{itemize}

A more detailed discussion of our post-hoc analysis can be found in the supplementary materials, section \ref{sec:posthoc}

\begin{figure}[tb]
    \centering
    \begin{subfigure}[t]{0.49\textwidth}
        \centering
        \includegraphics[width=\columnwidth]{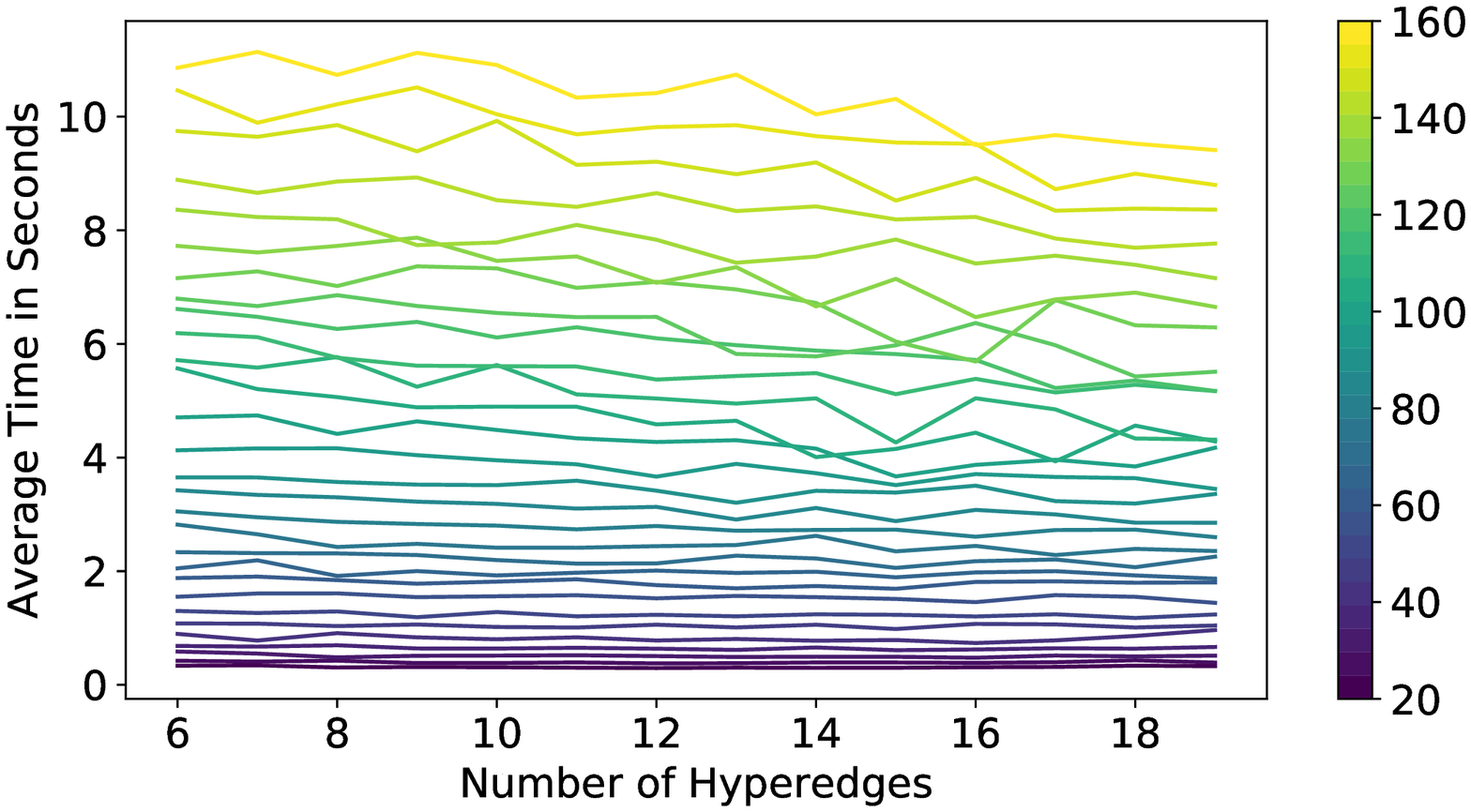}
        \caption{Line plot of median computation time by number of hyperedges. Lines are colored sequentially by the number of vertices, from 20 to 160 by 5's.}
    \end{subfigure}
    
    \begin{subfigure}[t]{0.49\textwidth}
        \centering
        \includegraphics[width=\columnwidth]{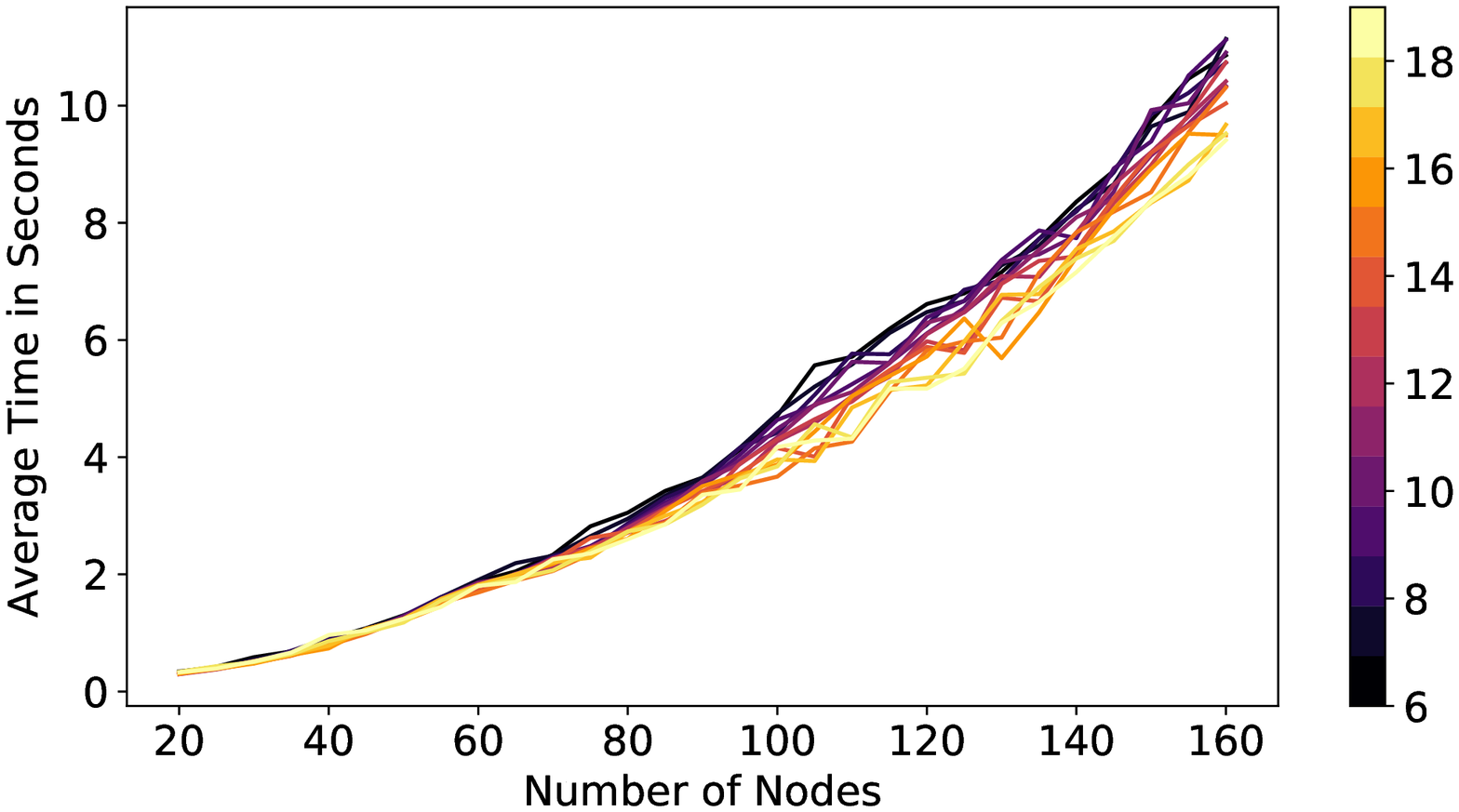}
        \caption{Line plot of median computation time by number of vertices. Lines are colored sequentially by the number of hyperedges, from 6 to 19. Note the much closer spread; the number of vertices is a very reliable predictor for running time.}
    \end{subfigure}
    \caption{Running times of the Balanced pipeline}
    \label{fig:time}
\end{figure}

\subsection{Scalability of the Default Pipeline}

Because the balanced pipeline will likely be the one used by the majority of MetroSets users, we subjected it to additional scrutiny, using it to make maps for every one of the 4060 datasets we extracted. By recording the time required to create each map, we were then able to use linear regression to create a model predicting the number of seconds it will take for a given dataset to be visualized using this preset pipeline. This model, with $R^2 = 0.99$, is: $T(N) = 4.07 \times 10^{-4} N^2$.

This implies a run-time that is quadratic in the number of vertices. In practical terms (backed by 4060 examples) the default pipeline can comfortably handle inputs of up to 140 vertices and 19 hyperedges in no more than 10 seconds; see \autoref{fig:time}.

\begin{figure*}[t]
    \centering
    \includegraphics[width=0.81\textwidth]{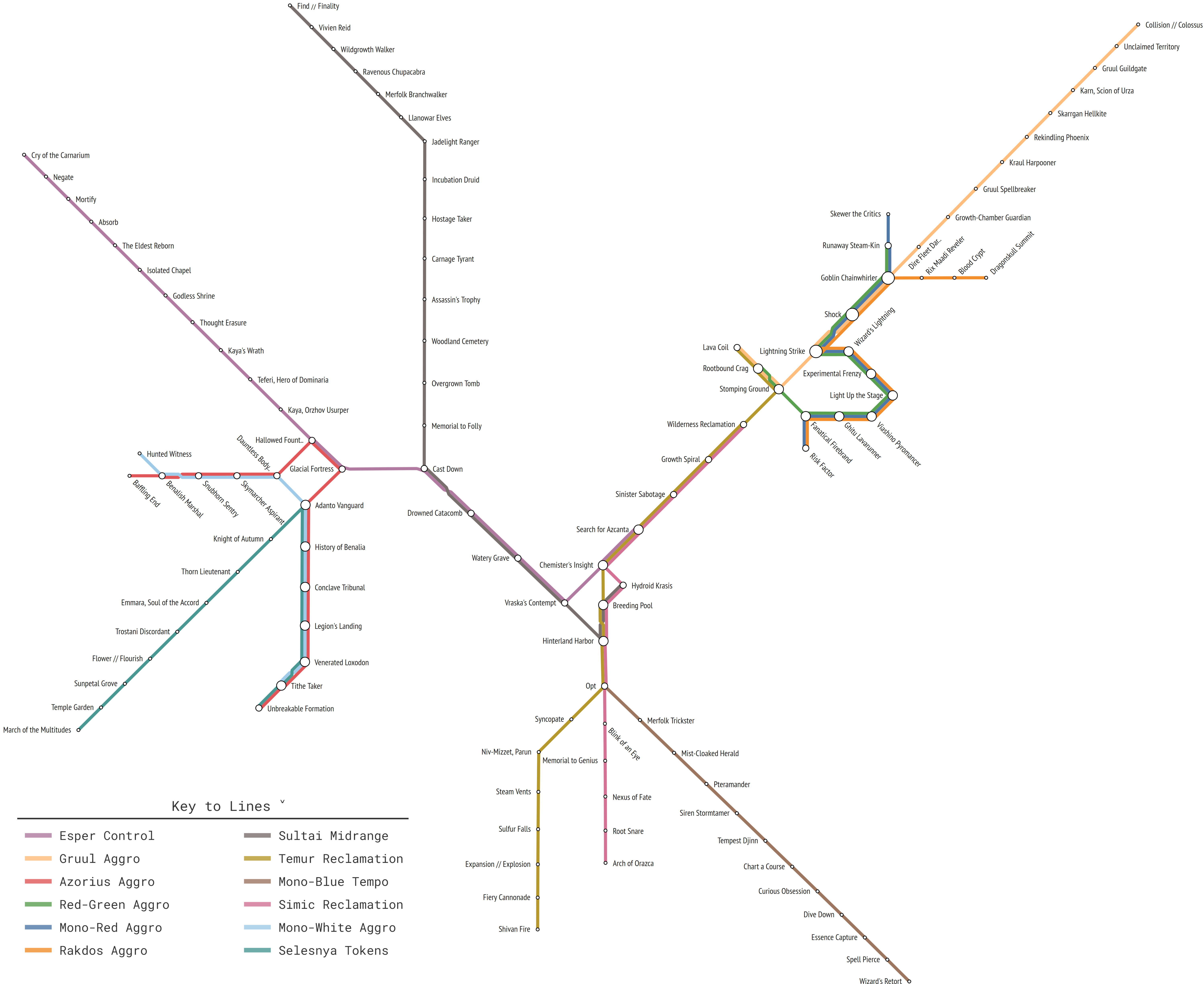}
    \caption{\new{``Magic: The Gathering" data set visualized with the MetroSets \textit{simplicity} pipeline.}}
    \label{fig:mtgPipeline}
\end{figure*}

\section{Discussion and Limitations}

\subsection{Comparison with Other Systems}

We created a metro map schematization of the Magic: The Gathering meta game data set, see \autoref{fig:mtgPipeline}, and directly compared our proposed pipeline with other set visualization systems. We choose this data set because it is a good representation of a medium sized data set. It has 12 sets and 112 elements. 67 of these elements belong to exactly one set. There are 3 elements that are members of 4 sets, which is the upper bound. Additionally, this hypergraph representation has a path-based planar support. We surveyed a collection of 6 tools that are concerned with similar set visualization tasks and could handle our data set without need for contextual information. A more thorough justification and evaluation, as well as the generated visualizations, can be found in Section~\ref{sec:magicComp} and \autoref{fig:comparsionTools} in the supplementary materials.

Three of the tools are from the category of Venn and Euler diagrams \cite{w-eaacved-12, larsson2020eulerr, rodgers2015visualizing}. The outputs are different from ours (for example, they do not explicitly show all elements).   The other three are matrix-based systems \cite{smds-ovtlbd-14, kls-vcwpmd-07, lamy2019rainbio}, which provide powerful capabilities to analyze data. The outputs again are different from ours (for example, there is no clear overview of the entire data) and interactions are required in order to perform most tasks.

We also compared a hand-made designer visualization~\cite{antoniazzi} to the MetroSets output for the same data. While the artist's map is visually pleasing and better utilizes available space, our map orders the vertices of the support graph better, which results in more conjoint lines; see \autoref{fig:comparisonArtist} in the supplementary materials for a side-by-side comparison.

Lastly, we evaluated a mixed-integer programming (MIP) approach for the schematization~\cite{nw-dlhqm-11}, which is known to be slow but of high visual quality for traditional metro maps, results in drastically better MetroSets layouts. Our findings suggest that while the MIP schematization is superior in regards to some quality measures ( octolinearity, monotonicity), it often fails to generate a schematization in reasonable time and is susceptible to edge crossings in the initial layout of Step 3; 
see \autoref{fig:mtgMip} in the supplementary materials for a comparison.   

\subsection{Limitations}

The MetroSets system is capable of visualizing a variety of different input data sets, but there are several limitations that require consideration. Due to the quadratic runtime, in the online setting the system is practically limited to inputs with no more that 200 vertices and 20 hyperedges. While we attempted to compare MetroSets to other techniques for visualizing set systems, we only managed to visualize our data with 6 of more than two dozen possible techniques (due to missing code, code that would not compile, difficulties entering the data, requirements for elements positioned in the plane, etc.). 
\new{A human subjects study evaluating MetroSets in comparison to alternative set visualization systems is an important next step, but beyond the scope of this paper.}
\new{We note that by its very nature the metro map metaphor might be affected by urban bias, and we do not know whether the effectiveness of the metaphor depends on familiarity with metro maps.}

Additional technical limitations include the lack of control over aspect ratio of the drawing area, poor outputs when any vertex must have degree greater than 8, and a limit of 20 on the number of hyperedges before line colors must be reused. \new{Space utilization is also not optimized, although adding a `compaction' step to the pipeline might be possible.} 
Currently MetroSets cannot handle multiple disconnected components or single-vertex sets. Although for reasonably sized inputs labels do not overlap, our system does not guarantee it. This could be accomplished by scaling the image until overlap is removed, but at the expense of larger drawing area. 
%

%
%

\section{Conclusions}
We presented MetroSets, a system for automatically visualizing set systems with hundreds of elements and more than a dozen intersecting sets using the metro map metaphor. We have designed MetroSets with as a flexible and modular 4-step layout pipeline and implemented a selection of algorithms for each step, optimizing various design goals. Our systematic quantitative evaluation explored the properties of these alternatives and their trade-offs. \new{We provide a functional prototype of MetroSets at \url{https://metrosets.ac.tuwien.ac.at} and have published the code on OSF \url{https://osf.io/nvd8e/}.}
\acknowledgments{
We thank the organizers of the Dagstuhl Seminar 17332 ``Scalable Set Visualizations'', specifically Robert Baker, Nan Cao, Yifan Hu, Michael Kaufmann, Tamara Mchedlidze, Sergey Pupyrev, Torsten Ueckerdt, and Alexander Wolff. We also thank Miranda Rintoul for her help  with the experimental analysis. This work is supported by NSF grants CCF-1740858, CCF-1712119, and
DMS-1839274 and by the Vienna Science and Technology Fund (WWTF) through project ICT19-035.}

\newpage
\bibliographystyle{abbrv-doi-hyperref}
\bibliography{metrosets}

\newpage
\appendix
\section{Supplemental Materials}

\subsection{The Two-Opt Heuristic}\label{sec:eval2opt}

It is not immediately obvious that the two-opt algorithm, which is traditionally applied in solving the metric travelling salesman problem, can be of use in solving the non-metric travelling salesman path problem. To quantify its usefulness, we performed a large number of experiments on random instances of travellings salesman problems resembling those likely to arise during the MetroSets experiments. On each of these instances, we compared results produced by the standard two-opt algorithm, a variation of the two-opt algorithm which constructs an initial tour using the nearest neighbor heuristic before refining it, and an ILP solver which finds the optimal solution.

\subsubsection{Experiment Design}

There are essentially three related versions of the travelling salesman problem which arise throughout the MetroSets pipeline. In the first case, which appears in the support graph construction, the cost of an edge between two vertices is based on their similarity score, which is generally the reciprocal of some small, positive integer. In the second case, which appears during the layout stage, the cost of an edge is the geometric mean of their similarity score and the Euclidean distance between the two vertices. In the third case, which appears on the final iteration of the layout stage, the cost is based purely on Euclidean distance. We tested the performance of the two-opt algorithm on each of these cases separately.

In all three cases, the experiment proceeds in essentially the same way. We consider inputs with between 3 and 60 vertices, inclusive (chosen to resemble the anticipated length of hyperedges visualized using our method). For each input size $n$, we randomly generate 10 instances of the travelling salesman problem by placing $n$ vertices uniformly throughout a square region with sides of 100 units. In the first two cases, we assign each pair of vertices a similarity score of 1 divided by a random integer from 1 to 5. Finally, we solve the travelling salesman path problem using the two-opt algorithm with arbitrary initial path, the two-opt algorithm with an initial path constructed using the nearest neighbor heuristic, and the ILP solver. The costs of the output paths are then compared.

\subsubsection{Results}

Across all three cases, the two-opt algorithm generally produced results no more than 10\% worse than the optimal solution. The use of the nearest neighbor heuristic for constructing the initial tour decreased this error by roughly half, while also speeding up the algorithm by a factor of roughly 5, presumably because it saves us the trouble of optimizing obviously terrible initial paths by a long and painful sequence of local improvements. On the basis of these results, which are summarized below, we conclude that the two-opt heuristic, with an initial path constructed using the nearest neighbor method, is an efficient and effective algorithm for our purposes.

\begin{table}[t]
	\centering
 \begin{tabular}{||>{\centering}p{2.8cm} >{\centering}p{1.4cm} >{\centering}p{1.4cm} >{\centering\arraybackslash}p{1.4cm} ||} 
 \hline
 Method & Avg. Cost & \% Above Optimal & Avg. Time (s) \\ [0.5ex]
 \hline\hline
 \multicolumn{4}{||c||}{Similarity Only}\\
 \hline
 Two-Opt & 6.24 & 3.03\% & .017 \\ 
 Two-Opt (NN Init) & 6.20 & 1.54\% & .003 \\
 Gurobi & 6.14 & 0\% & .019 \\ [1ex] 
 \hline
  \multicolumn{4}{||c||}{Hybrid}\\
 \hline
 Two-Opt & 56.12 & 5.18\% & .026 \\ 
 Two-Opt (NN Init) & 55.33 & 3.09\% & .006 \\
 Gurobi & 53.52 & 0\% & .039 \\ [1ex] 
 \hline
  \multicolumn{4}{||c||}{Euclidean Only}\\
 \hline
 Two-Opt & 431.92 & 8.36\% & .026 \\ 
 Two-Opt (NN Init) & 415.64 & 4.06\% & .005 \\
 Gurobi & 398.35 & 0\% & .055 \\ [1ex] 
 \hline
\end{tabular}
\caption{ Summary of experimental results evaluating quality of the two-opt heuristic.} \label{fig:two-opt-table}
\end{table}

\subsection{Comparison}

\subsubsection{Set visualization tasks}

We used the proposed task taxonomy of  Alsallakh et al.~\cite{alsallakh2014visualizing} and self-evaluated our pipeline. \autoref{tab:interactivity} shows the tasks of the three different categories and how they are performed in MetroSets.

\begin{table*}[t]
\resizebox{\textwidth}{!}{%
\begin{tabular}{|c|l|c|l|}
\hline
\textbf{\#} & \multicolumn{1}{c|}{\textbf{Task}}                                                 & \textbf{Supported} & \multicolumn{1}{c|}{\textbf{How to perform the task}} \\ \hline
A1        & Find/Select elements that belong to a specific set                                 & V, Id                & Hover line to highlight elements                      \\
A2        & Find sets containing a specific element.                                           & V, Id                & Hover element to highlight sets                       \\
A3        & Find/Select elements based on their set memberships                                & V                & Check lines for elements                             \\
A4        & Find/Select elements in a set with a specific set member-ship degree               & V                & Count number of outgoing lines                        \\
A5        & Filter out elements based on their set memberships.                                & Io, Id                & Exclusive or Union mode and hover can act as filter                                                        \\
A6        & Filter  out  elements  based  on  their  set  membership  degrees                  & -                 &                                                       \\
A7        & Create a new set that contains certain elements.                                   & -                 &                                                       \\
\hline
B1        & Find out the number of sets in the set family.                                      & V, Id                & Information given by legend or sidebar                \\
B2        & Analyze inclusion relations.                                                      & V                & Conjoint metro lines depict inclusion                 \\
B3        & Analyze  inclusion  hierarchies                                                    & -                 & -                                                      \\
B4        & Analyze exclusion relation                                                         & V, Io                & Exclusion mode can check for exclusion             \\
B5        & Analyze intersection relation                                                      & Io                & Intersection mode can check for intersection          \\
B6        & Identify intersections between k sets                                              & V, Io                & Intersection mode can check for k intersections       \\
B7        & Identify the sets involved in a certain intersection                               & V, Id                & Hover lines or vertices to highlight intersecting sets    \\
B8        & Identify set intersections belonging to a specific set                             & Io                & Use intersection mode to check against other sets     \\
B9        & Identify the set with the largest / smallest number of pair-wise set intersections & Io                 & Use intersection mode to identify                                                       \\
B10       & Analyze and compare set- and intersection cardinalities                            & V, Io                & Use intersection mode to check cardinality            \\
B11       & Analyze and compare set similarities                                               & -                 & -                                                      \\
B12       & Analyze and compare set exclusiveness                                               & Io                & Exclusion mode or can show set exclusiveness                                                      \\
B13       & Highlight  specific  sets,  subsets,  or  set  relations                           & V, Id                & Hovering or clicking emphasizes sets                  \\
B14       & Create a new set using set-theoretic operation                                     & -                 & -                                                      \\
\hline
C1        & Find out the attribute values of a certain element                                 & Id                & Hover element to see attributes                       \\
C2        & Find out the distribution of an attribute in a certain set or subset                & -                & -                                                      \\
C3        & Compare the attribute values between two sets or subsets                           & -                 & -   
\\
C4        & Analyze  the  set  memberships  for  elements  having  certain attribute values    & -                 & -   
\\
C5        & Create  a  new  set  out  of  elements  that  have  certain  attribute values      & -                 & -                                                      \\ \hline
\end{tabular}%
}
\caption{We used the task taxonomy of Alsallakh et al. to classify our proposed system. We state how a certain task can be solved and whether interactivity is required or not. (\textit{V}: Task can be solved visually, \textit{Id}: Details-on-demand with interactivity is necessary to solve the task, \textit{Io}: Interactivity with the set operation mode is required, \textit{-}: Task is not supported)}  
\label{tab:interactivity}
\end{table*}

\subsubsection{Comparison with Other Systems} \label{sec:magicComp}

We created a metro map schematization of the Magic: The Gathering meta game data set and directly compared our proposed pipeline with other set visualization systems.
Magic: The Gathering is a collectible card game that is played by two players against each other with their own 60 card decks. The meta game is the subset of successful decks the players gravitate towards when entering a tournament and choosing amongst thousands of legal cards. The data set contains the top 12 decks as of spring 2019 and each meta deck is modeled as a hyperedge, depicting cards as vertices in said hyperedges. Some cards may appear in several decks because of their usefulness. 

\begin{figure*}[p]
    \begin{subfigure}[t!]{\columnwidth}
      \begin{subfigure}[t]{0.95\columnwidth}
        \includegraphics[width=\linewidth, keepaspectratio]{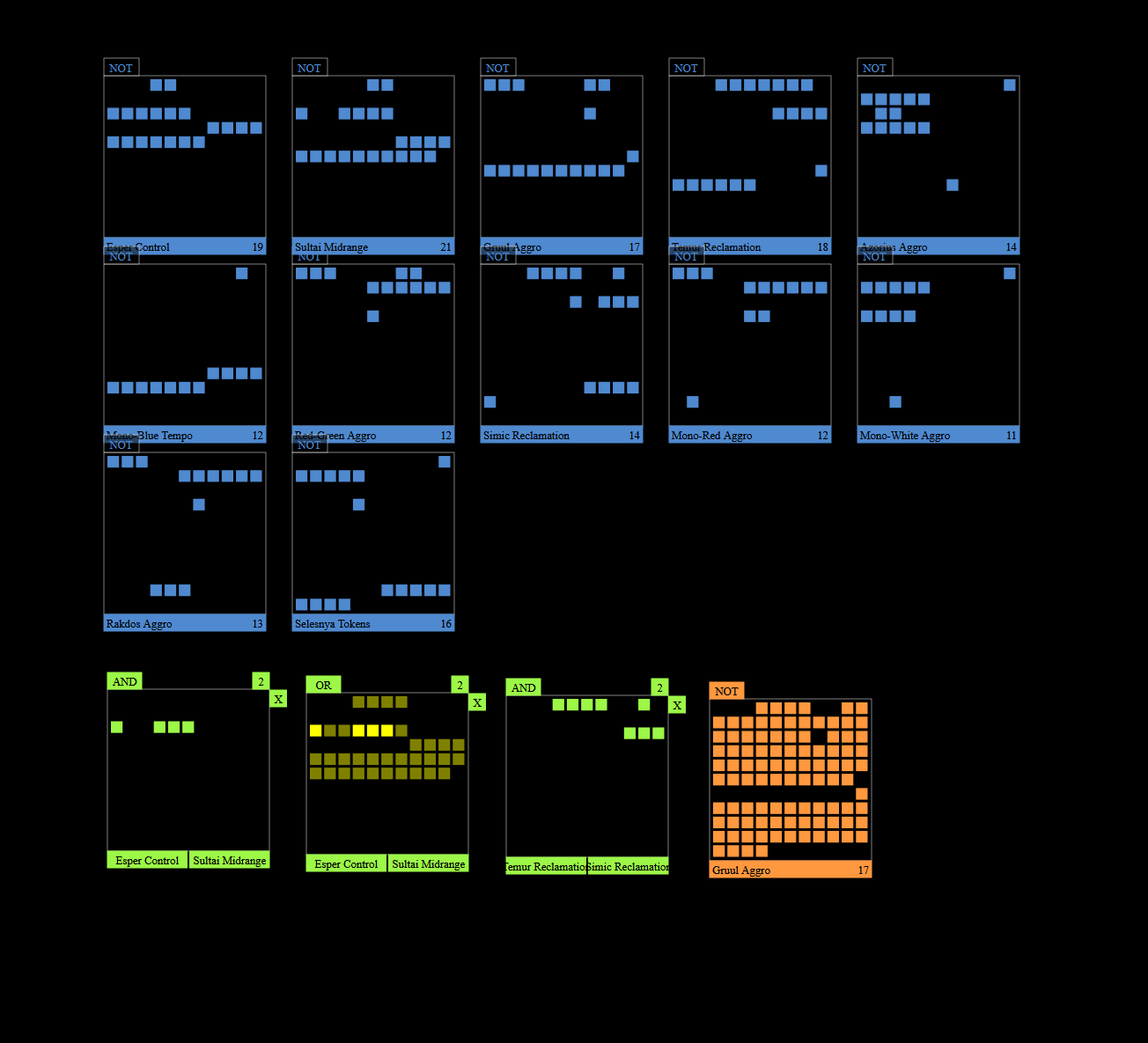}
        \subcaption{OnSet \cite{smds-ovtlbd-14} visualizes sets as individual matrices. Additional information about the data, e.g. number of set elements, is shown next to the boxes. Set elements are represented as colored squares with a unique position. Dragging and dropping sets over one other performs set operations, like union or intersection. Performing such operations are helpful to gain more insight, but it is impossible without interaction to see which element is which square.}
      \end{subfigure}
      \par\bigskip
      \begin{subfigure}[t]{0.95\columnwidth}
        \includegraphics[width=\linewidth, keepaspectratio]{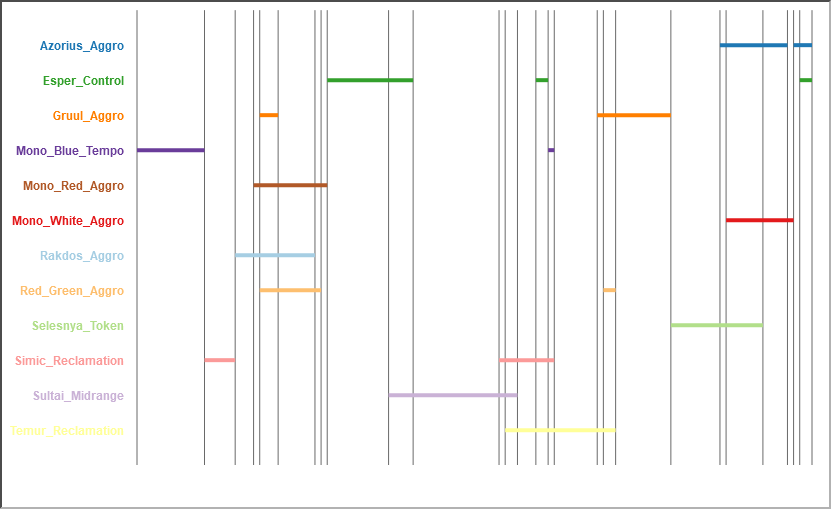}
        \subcaption{LinearDiagrams \cite{rodgers2015visualizing} visualizes sets as horizontal lines and membership as vertical bars. It is classified among the Euler-based set diagrams by the characteristic that it generates area-proportional diagrams. The size of the vertical bars is proportionate to the number of elements in their sets and their intersections. There is no information given on individual set elements.}
      \end{subfigure}
      \par\bigskip
      \begin{subfigure}[b]{0.95\columnwidth}
        \includegraphics[width=\linewidth, keepaspectratio]{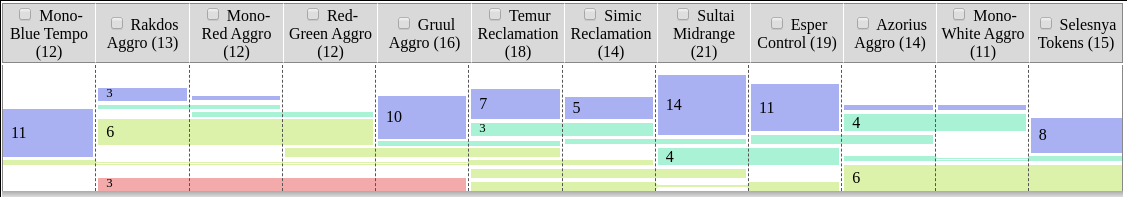}
        \subcaption{RainBio \cite{lamy2019rainbio} is a recent tool designed for visualizing large sets biology. It represents each set as a column, while elements which belong to the exclusive intersection of 1 or more sets are drawn as colored boxes which span the columns of those sets. Boxes with more than a small number of elements are labeled with their cardinality, and with user interaction it is possible to see all of the elements contained in each box or to filter out intersections with cardinality below a certain threshold.}
      \end{subfigure}
    \end{subfigure}
    \hfill
    \begin{subfigure}[t!]{\columnwidth}
      \begin{subfigure}[t!]{0.95\columnwidth}
        \centering
        \includegraphics[width=\linewidth, keepaspectratio]{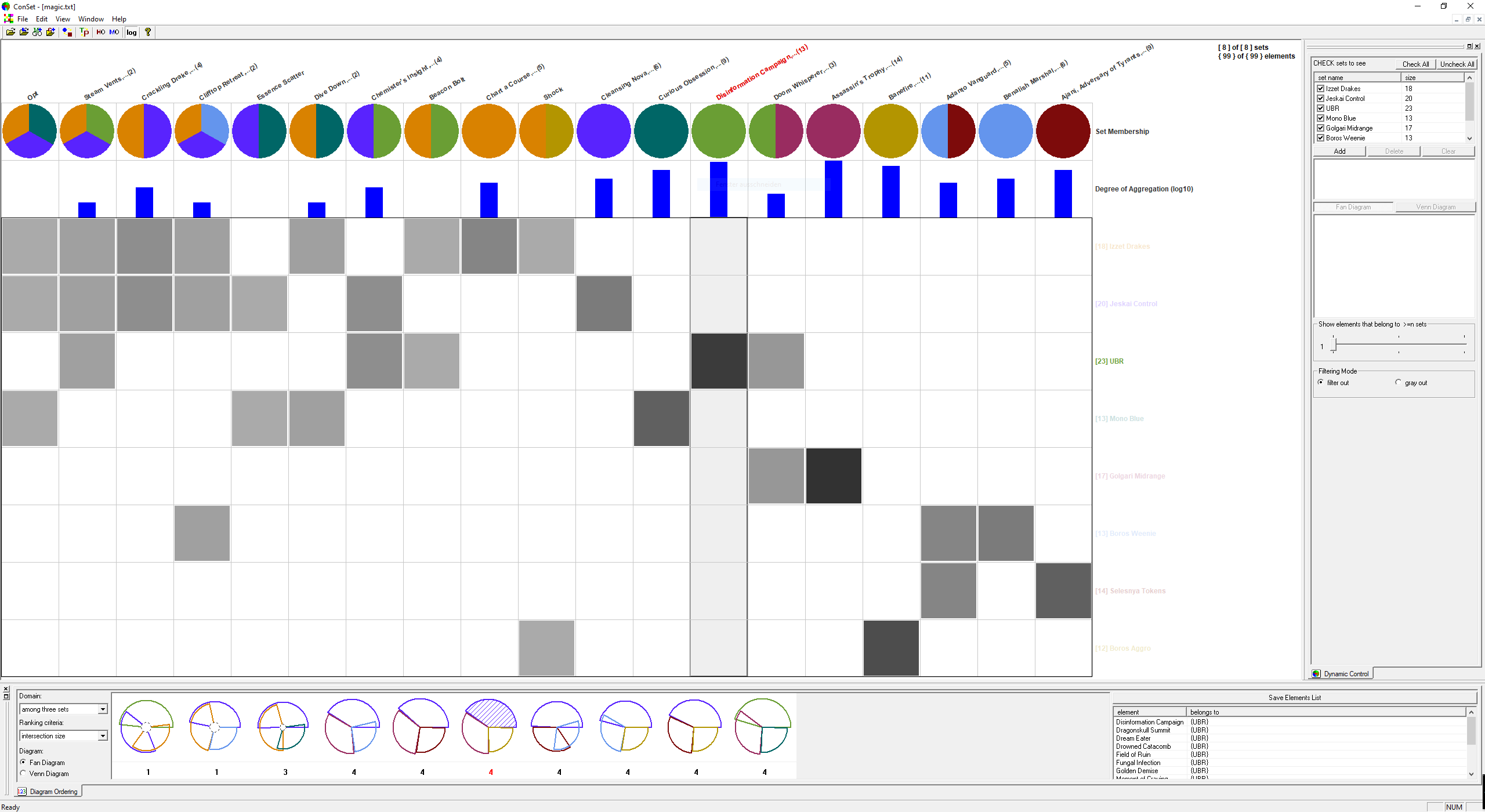}
        \subcaption{ConSet \cite{kls-vcwpmd-07} uses a matrix as metaphor for visualizing set membership. Sets are mapped as rows and elements as columns. Multiple elements that share the same set membership are aggregated. Cells represent the set membership and sequential coloring is used to map the total number of elements per set. The columns and rows of the matrix are reorderable to highlight different aspects of the data. Venn diagrams are displayed at the bottom that show set intersection. The tool also provides a tree view, that is used for visualizing individual set elements. It is also possible to filter the data. We found that ConSet does not display the elements that are members of 4 sets.}
      \end{subfigure}
      \par\bigskip
      \begin{subfigure}[t!]{0.95\columnwidth}
        \centering
        \includegraphics[width=0.8\linewidth, keepaspectratio]{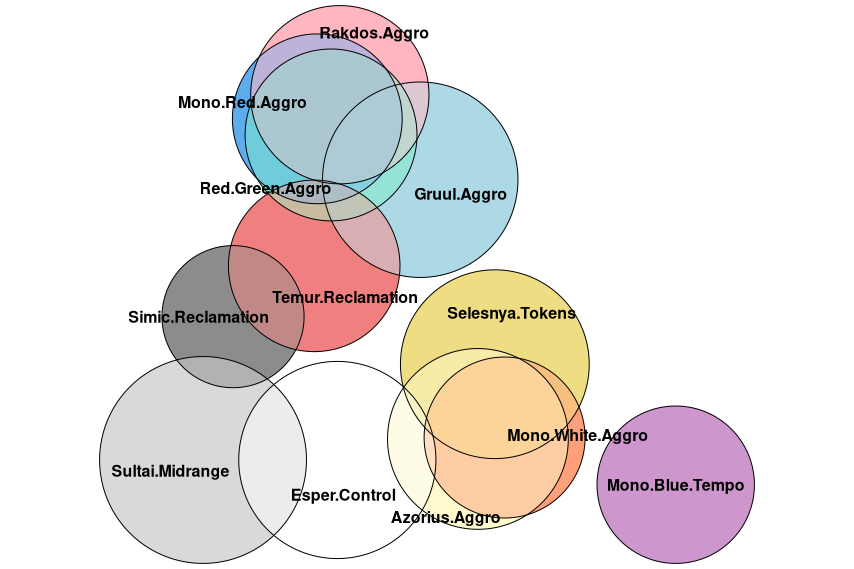}
        \subcaption{eulerr \cite{larsson2020eulerr} is a popular R package for automatically generating area proportional Euler diagrams. The images it provides are static, and it is not possible to view individual elements. In addition, in attempting to minimize error in area proportionality, it can produce misleading results. In this example, Mono Blue Tempo is incorrectly shown as disjoint, when in fact it shares one element with Simic Reclamation and Temur Reclamation. }
      \end{subfigure}
      \par\bigskip
      \begin{subfigure}[t!]{0.95\columnwidth}
        \centering
        \includegraphics[width=0.6\linewidth, keepaspectratio]{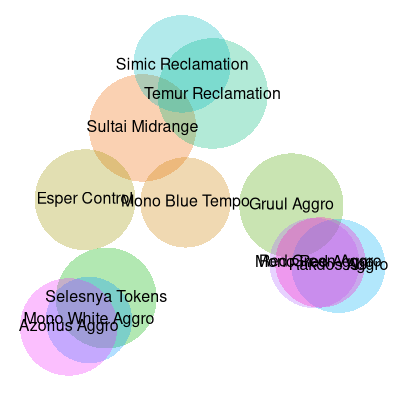}
        \subcaption{venneuler \cite{w-eaacved-12} is a second, somewhat older R package for generating area proportional Euler diagrams. As with eulerr, it produces static images which do not give information on individual elements. Also like eulerr, it can occasionally produce misleading images, as in this case where Temur Reclamation falsely appears to have exclusive overlap with Sultai Midrange.}
      \end{subfigure}
    \end{subfigure}
  \caption{Comparison of how various tools handle the Magic: The Gathering data set.}
  \label{fig:comparsionTools}
\end{figure*}

First we excluded set visualization systems which were not sensible for comparison against our proposed system, because of their differences in prerequirements for the input data. These are mainly line-based, region-based or glyph-based overlay systems, like Bubble Sets, LineSets or Kelp Diagrams , where an initial position for all set elements is required. Therefore, we think that comparing those systems would be too dependent on the choice of initial positions.  
We also excluded tools that would have been interesting for visualizing our data set, but we were not able to find implementations, the implementations did not work on our end, or they were not flexible enough to allow user data.
Finally we surveyed 6 tools and generated an output for each. All tools and their respective output are depicted in \autoref{fig:comparsionTools}.

\begin{figure*}[p]
    \centering
    \includegraphics[width=\textwidth]{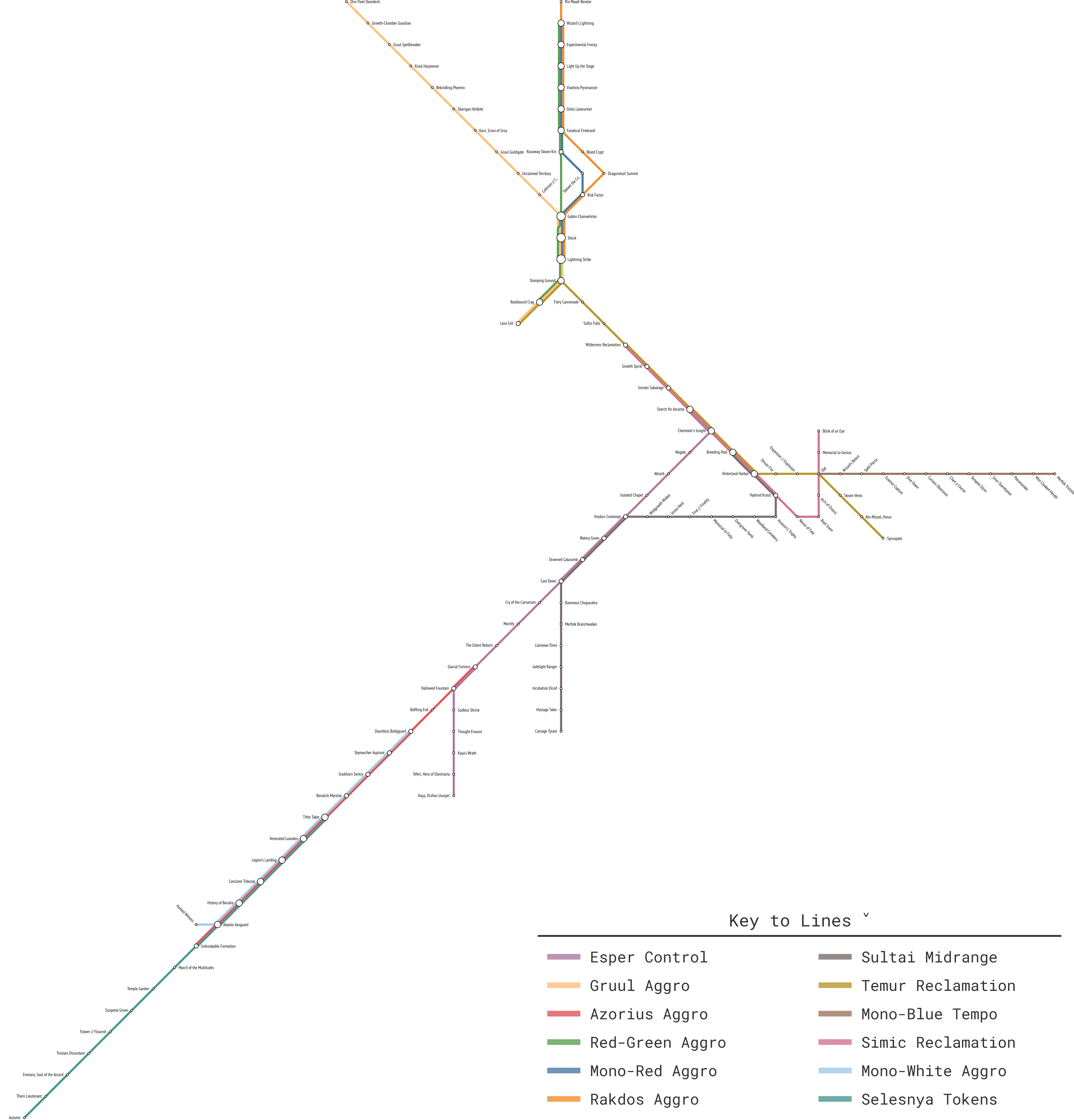}
    \caption{The Magic: The Gathering data set visualized with our pipeline. We used a mixed integer programming schematization \cite{nw-dlhqm-11} to schematize the initial layout, instead of one of the two pipeline options.}
    \label{fig:mtgMip}
\end{figure*}

\begin{figure*}[t]
  \centering
  \begin{subfigure}[b!]{0.55\textwidth}
    \centering
    \includegraphics[width=\linewidth, keepaspectratio]{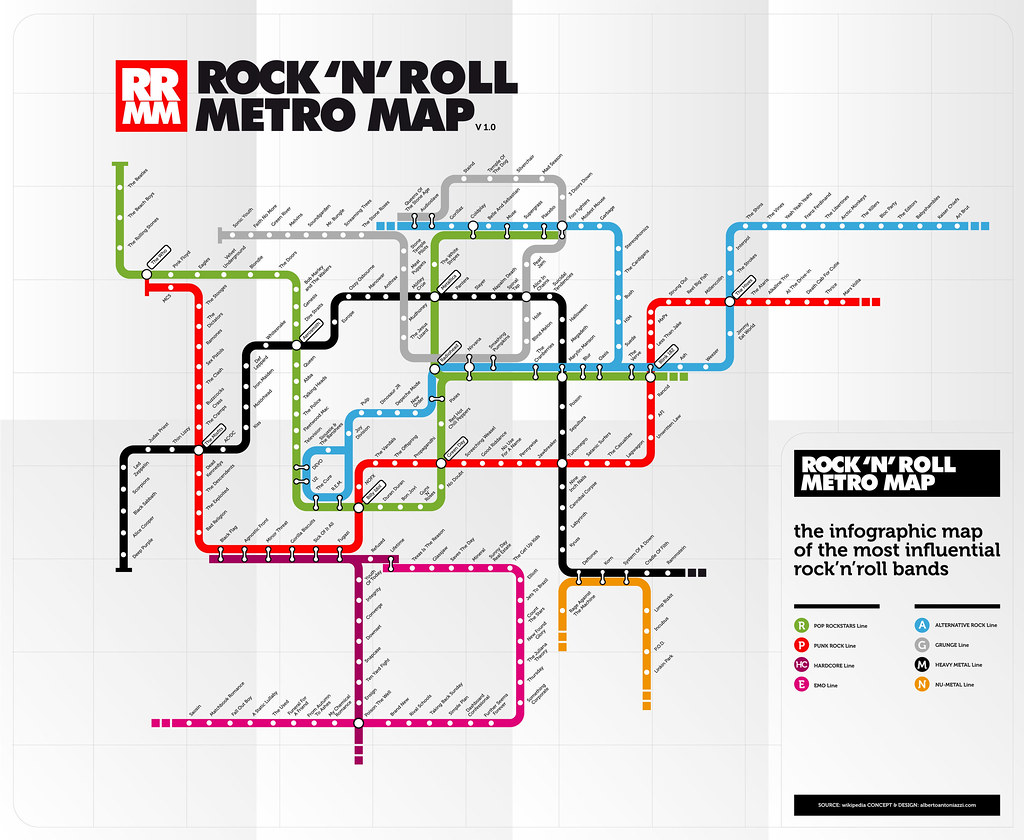}
    \subcaption{The Rock'n'Roll metro map by designer Alberto Antoniazzi \cite{antoniazzi}. His design only uses horizontal and vertical lines while maximizing the utilization of the available space.}
  \end{subfigure}
  \par \bigskip
  \begin{subfigure}[b!]{0.95\textwidth}
    \centering
    \includegraphics[width=0.8\linewidth, keepaspectratio]{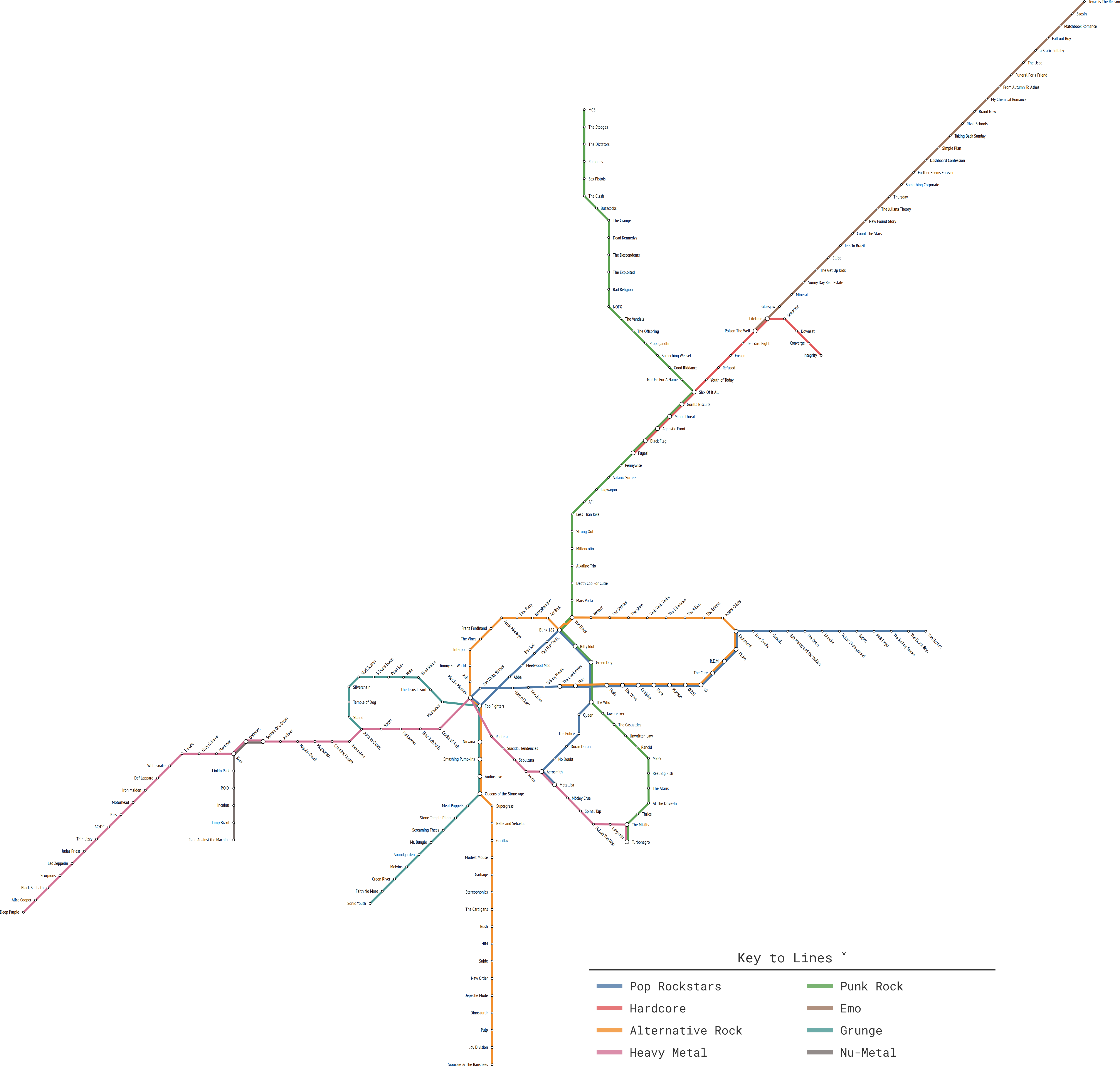}
    \subcaption{Schematization created with our pipeline.}
  \end{subfigure}
  \caption{Comparison between manually created schematization by an artist and our pipeline. The data set used is the Rock'n'Roll data set, which has 226 elements depicting bands and 8 sets depicting music genres.}
  \label{fig:comparisonArtist}
\end{figure*}

\subsection{Post-Hoc Analysis of Experiment Results}\label{sec:posthoc}

\textit{Running time} grows fairly slowly as the number of vertices approaches 100, and grows very quickly thereafter. The number of hyperedges, on the other hand, has relatively little impact.

The pipeline step with the greatest impact is the layout stage. TSP (Neato) is slightly faster than TSP (Kamada-Kawai) (p = .022, mean difference = 1.01s), which is in turn much faster than the Spring Embedder (p = .001, mean difference = 6.15s). This effect is exacerbated as the number of vertices in the input graph increases; for graphs with 40 vertices, the Spring Embedder is not significantly slower than the other layout algorithms. However, at 160 vertices, it averages 13.9 seconds slower than TSP (Kamada-Kawai) and 16.2 seconds slower than TSP (Neato).

With respect to the other pipeline stages, the Consecutive Ones method of support graph construction is slightly slower than the TSP method (p = .001, mean difference = .88s), and Chivers-Rodgers is slightly slower than Least Squares Approximation for schematization (p = .001, mean difference = 1.23).

Maximum error in \textit{Octolinearity} is minimized by choosing Chivers-Rodgers for schematization (p = .001, mean difference of 7.3 degrees). Chivers-Rodgers is also slightly superior at minimizing average error (p = .001, mean difference of .62 degrees). 

Interestingly, adding vertices to a dataset has a beneficial effect on average octolinearity. This is likely because low degree vertices can frequently be drawn in a perfectly octolinear fashion, and thus tend to drag down the average.

The two TSP algorithms are not significantly different with respect to \textit{Average Edge Uniformity} (p=.4), but both are significantly better than the Spring Embedder (p=.001, mean difference of .035 for TSP (Neato) and .043 for TSP (Kamada-Kawai)). The same pattern holds with maximum error in edge uniformity; the TSP algorithms do not perform significantly differently (p=.43), but outperform the Spring Embedder (p=.001, mean difference of .22 for TSP (Neato) and .26 for TSP (Kamada-Kawai)).

With respect to \textit{Monotonicity}, the two TSP algorithms are not significantly different (p=.51), but both outperform the Spring Embedder (p=.001, mean difference of 2.8 for TSP (Neato) and 2.4 for TSP (Kamada-Kawai)).

The same holds for \textit{Gabriel Score}: the TSP layout algorithms perform comparably (p=.84) but the Spring Embedder underperforms compared to both of them (p=.001, mean difference of 13.6 for TSP (Neato) and 14.5 for TSP (Kamada-Kawai)).

\textit{Consecutive Ones} exhibits several interesting properties. Firstly, there is no significant difference in the Consecutive Ones property for number of hyperedges between 6 and 15; however, the score deteriorates significantly for graphs with 18 hyperedges (p=.001, mean difference of roughly 1.3 with each of the smaller values). Secondly, like average error in octolinearity, the consecutive ones property is improved by the addition of more vertices, with the worst performance by far seen in graphs with 40 vertices (p=.001, mean difference of roughly 0.8 compared to larger values).

While it is not statistically significant, it is also interesting to note that, while the Spring Embedder was the worst performing algorithm when paired with the TSP method for support graph construction, it was the best performing algorithm when paired with the consecutive ones method. This possibly hints at why the support graph choice has a surprisingly low impact; during both of the TSP layout algorithms, the paths representing each hyperedge are iteratively changed according to optimization metrics mostly independent of the consecutive ones property, meaning that some of the work done by the support graph algorithm is erased. The spring embedder preserves the paths given to it, and while this can lead to poor performance with respect to other metrics, it also means that does the best job of preserving the optimization of the support graph algorithm.

The number of \textit{Edge Crossings} does not differ significantly between the two TSP layout algorithms, (p = 0.2), but the Spring Embedder performed worse than both of them (p = .001, mean difference of 3.7 for TSP (Neato) and 4.7 for TSP (Kamada-Kawai)).

The number of \textit{Self Crossings} is dominated by the choice of layout algorithm. This is almost certainly a result of the fact that the two-opt heuristic for solving the Travelling Salesman Path Problem removes almost all self-crossings. Accordingly, both TSP algorithms reliably produce maps with 0 self-crossings, with a few very rare exceptions. The Spring Embedder makes no such guarantee, however. While it is not significantly worse than the TSP algorithms for graphs with 6 hyperedges, it averages over 2 self-crossings for graphs with 18 hyperedges.

The number of \textit{Line Crossings} exhibits similar behavior to the Consecutive Ones property: although the Spring Embedder is the worst performing layout algorithm when paired with the TSP support graph algorithm, it is the best performing algorithm when paired with the Consecutive Ones support graph algorithm. This time, however, the differences are mostly statistically significant, with $p \leq .01$ for every other combination of support and layout algorithm except for Consecutive Ones + TSP (Neato), for which $p = .07$. Mean differences range from .7 to 1.5.

\begin{table*}
	\centering
 \begin{tabularx}{\textwidth}{|| c | YYYY ||}
 \hline
 & \multicolumn{4}{c||}{Choose a pipeline with:} \\
 To optimize... & Support & Insertion & Layout & Schematization \\
 \hline
 Running Time & Travelling Salesman & * & TSP (Neato) & Least Squares \\
 Avg. Oct  & Travelling Salesman & * & TSP (Kamada-Kawai) & Chivers-Rodgers \\
 Max Oct & * & Split Insert & TSP (Kamada-Kawai) & Chivers-Rodgers \\
 Avg. Edge Uni & * & * & TSP (*) & Chivers-Rodgers \\
 Max Edge Uni & * & * & TSP(*) & Chivers-Rodgers \\
 Monotonicity & Travelling Salesman & First Viable & TSP(*) & Chivers-Rodgers \\
 Gabriel Score & Travelling Salesman & First Viable & TSP(*) & * \\
 Con. Ones & Consecutive Ones & * & Spring Embedder & * \\
 Edge Crossings & Travelling Salesman & * & TSP (Kamada-Kawai) & * \\
 Self Crossings & * & Split Insert & TSP(*) & * \\
 Line Crossings & Consecutive Ones & * & Spring Embedder & * \\
 \hline
\end{tabularx}
\caption{ Suggested pipelines for specific optimization goals, based on post-hoc analysis of our experimental results. * denotes that any choice of pipeline step is equally acceptable.}
\label{tab:recommended-pipelines}
\end{table*}

\begin{figure*}[p]
    \begin{subfigure}[t!]{\columnwidth}
      \begin{subfigure}[t]{0.95\columnwidth}
        \includegraphics[width=\linewidth, keepaspectratio]{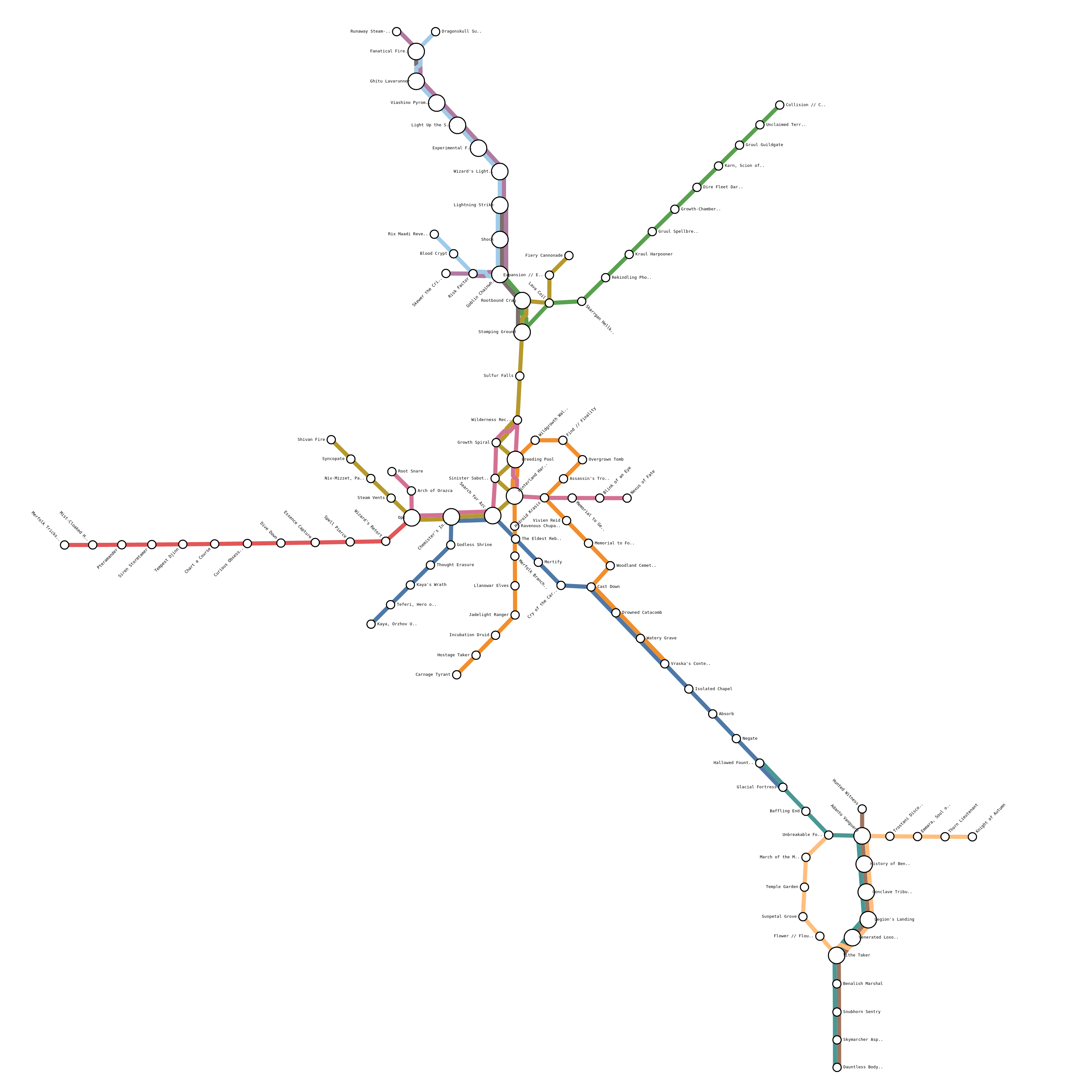}
        \subcaption{A visualization which scores poorly on Monotonicity. Notice the zig-zagging lines in the central area. The Monotonicity score of this map is 20.}
      \end{subfigure}
      \par\bigskip
      \begin{subfigure}[t]{0.95\columnwidth}
        \includegraphics[width=\linewidth, keepaspectratio]{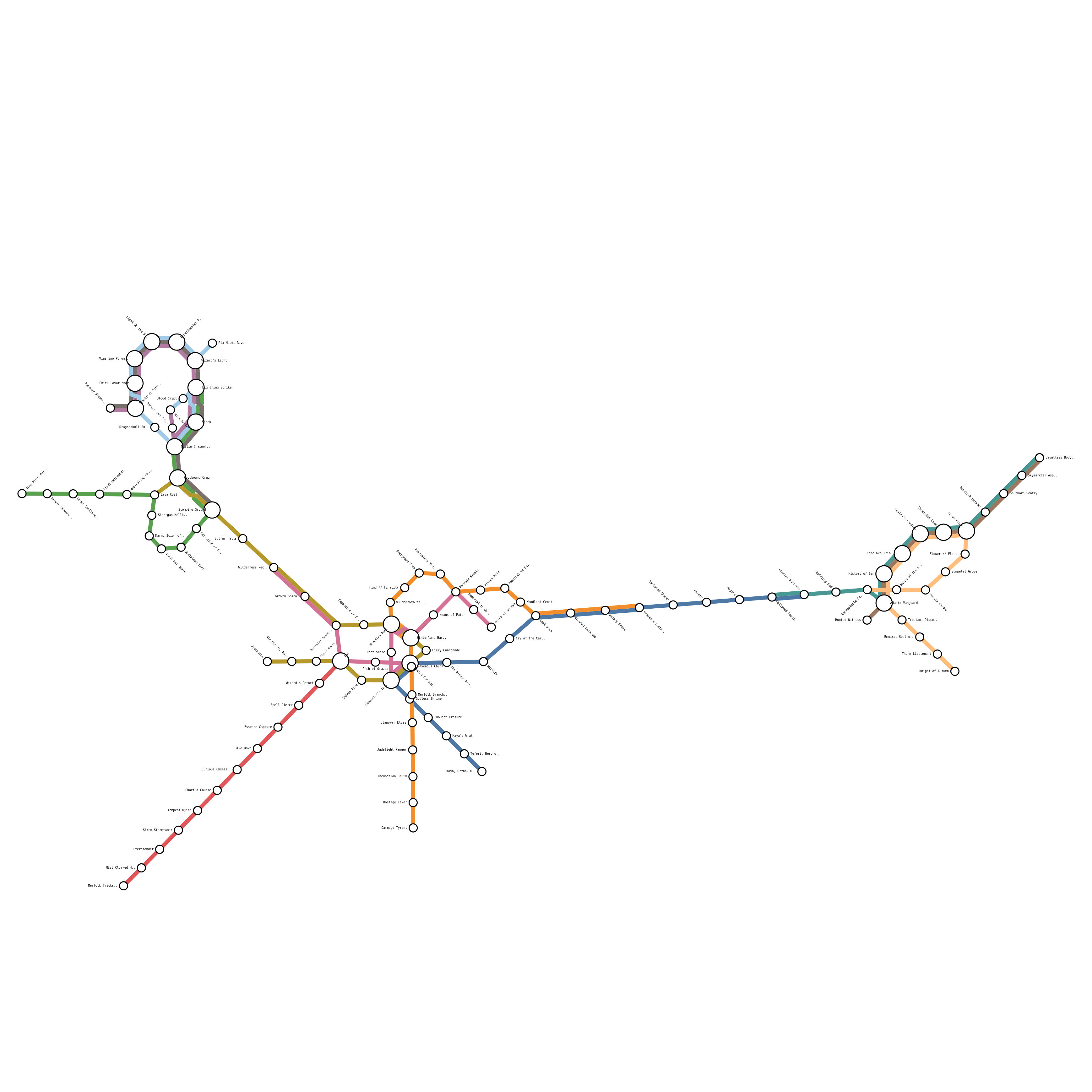}
        \subcaption{This drawing scores quite poorly on octolinearity, with an average error of roughly 2.2 degrees.}
      \end{subfigure}
    \end{subfigure}
    \hfill
    \begin{subfigure}[t!]{\columnwidth}
      \begin{subfigure}[t]{0.95\columnwidth}
        \includegraphics[width=\linewidth, keepaspectratio]{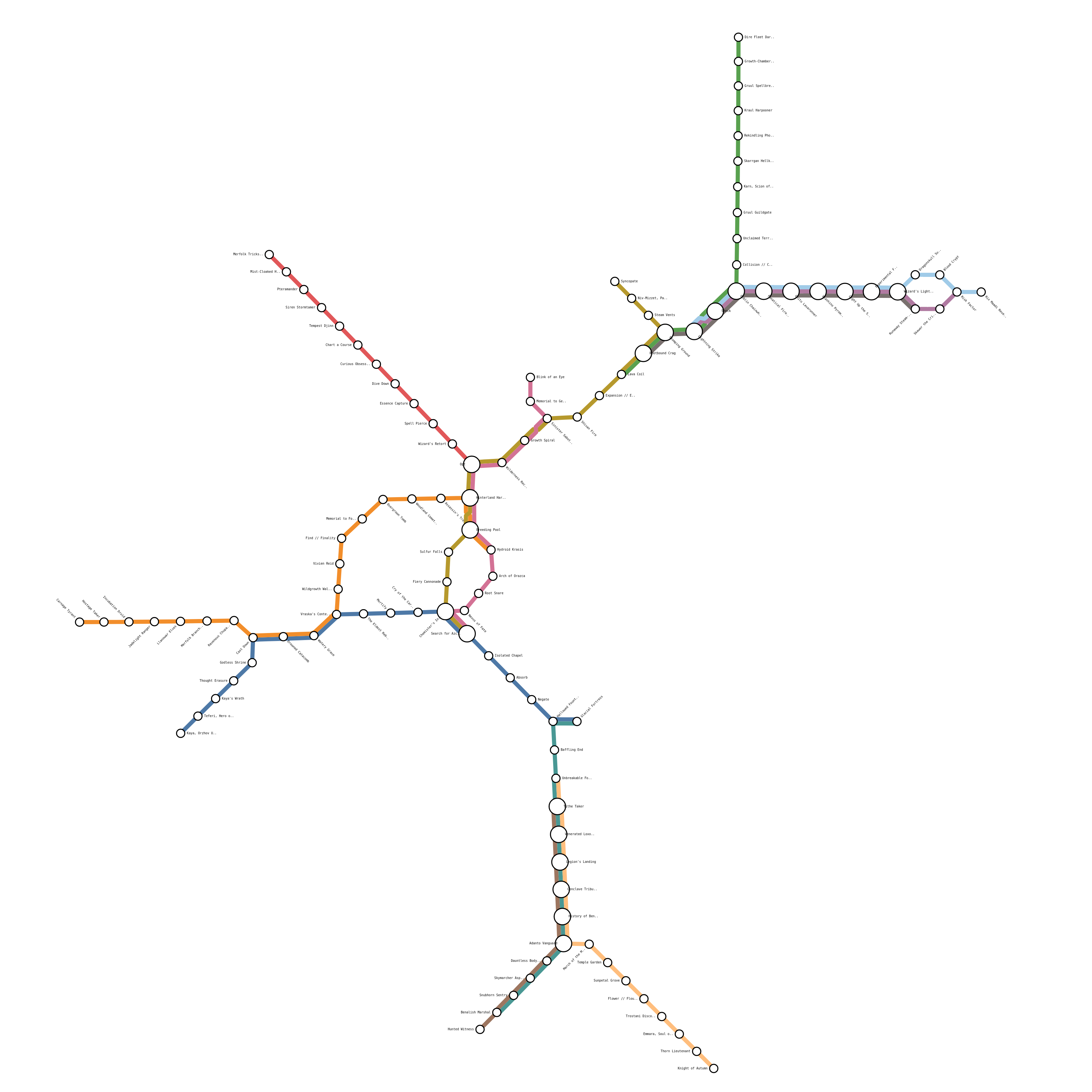}
        \subcaption{This is a visualization of the same dataset which scores much better on Monotonicity, and comparably on all other metrics. Its Monotonicity score is only 2.}
      \end{subfigure}
      \par\bigskip
      \begin{subfigure}[t]{0.95\columnwidth}
        \includegraphics[width=\linewidth, keepaspectratio]{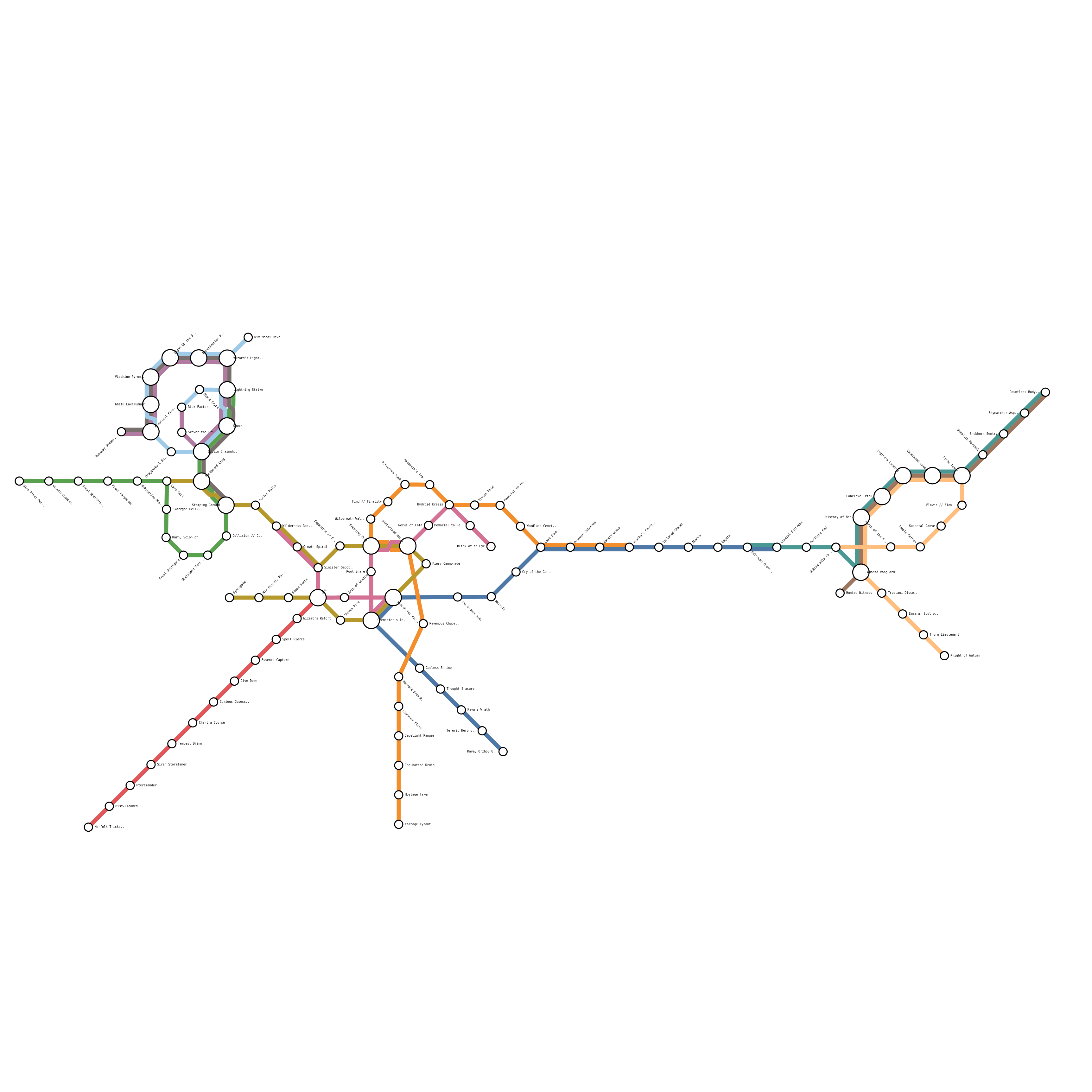}
        \subcaption{This visualization is structurally almost identical, but scores much better on octolinearity, with an average error of only .43 degrees.}
      \end{subfigure}
    \end{subfigure}
    \caption{These figures illustrate the Monotonicity and Octolinearity metrics using maps generated by our system. Each pair of maps scores very similarly on most metrics, but very differently on the metric in question. We believe that these images demonstrate that our chosen metrics are accurate measures of metro map quality (figure continues onto next page).}
    \label{fig:metriccomparison}
\end{figure*}

\begin{figure*}[tbp]
    \begin{centering}
    \begin{subfigure}[t!]{\columnwidth}
        \includegraphics[width=0.95\columnwidth, keepaspectratio]{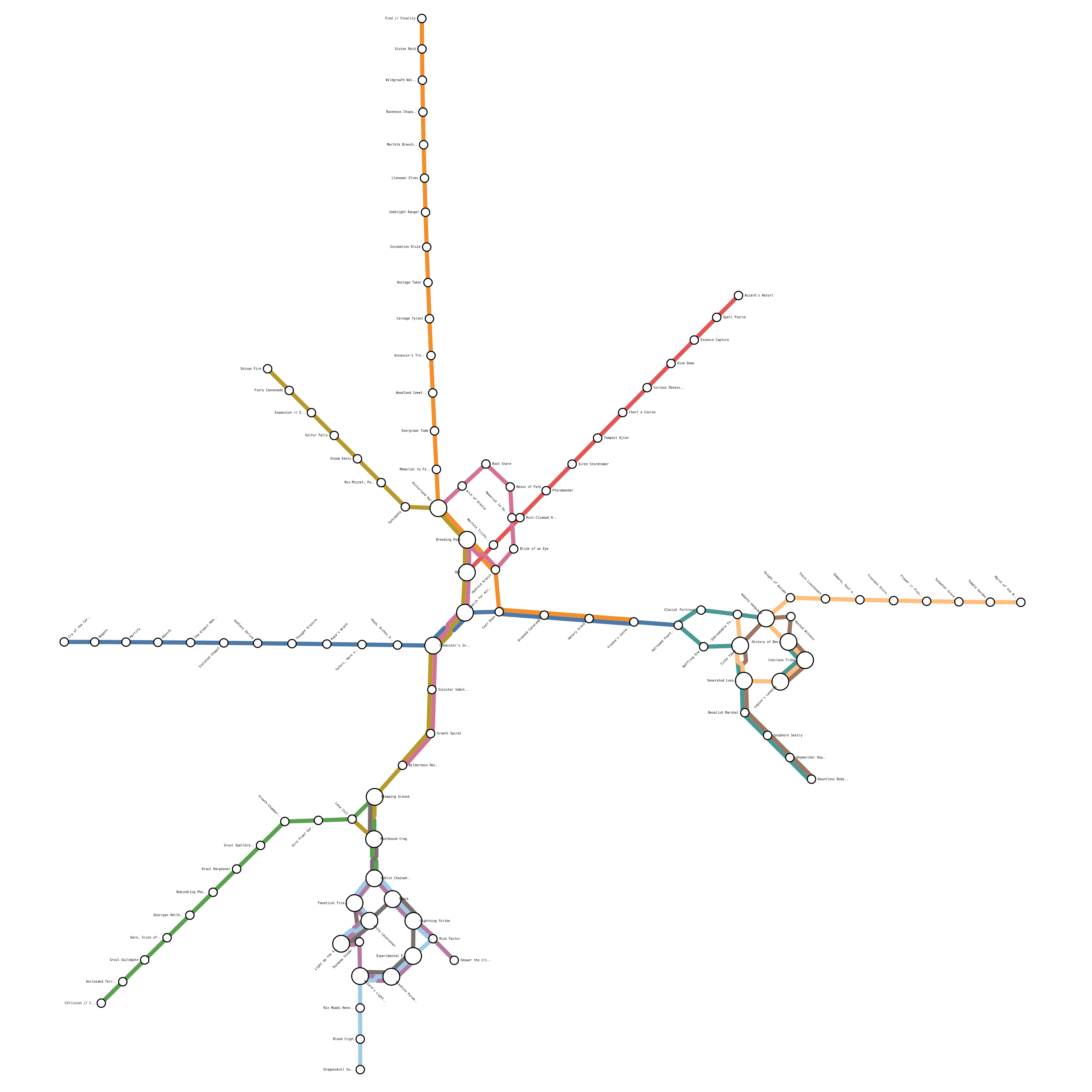}
        \subcaption{This map performs poorly by the Consecutive Ones metric, with a score of 11. Notice how the lines frequently split apart and rejoin at the bottom and righthand sides of the map.}
        
    \end{subfigure}
    \hfill
    \begin{subfigure}[t!]{\columnwidth}
        \includegraphics[width=0.95\columnwidth, keepaspectratio]{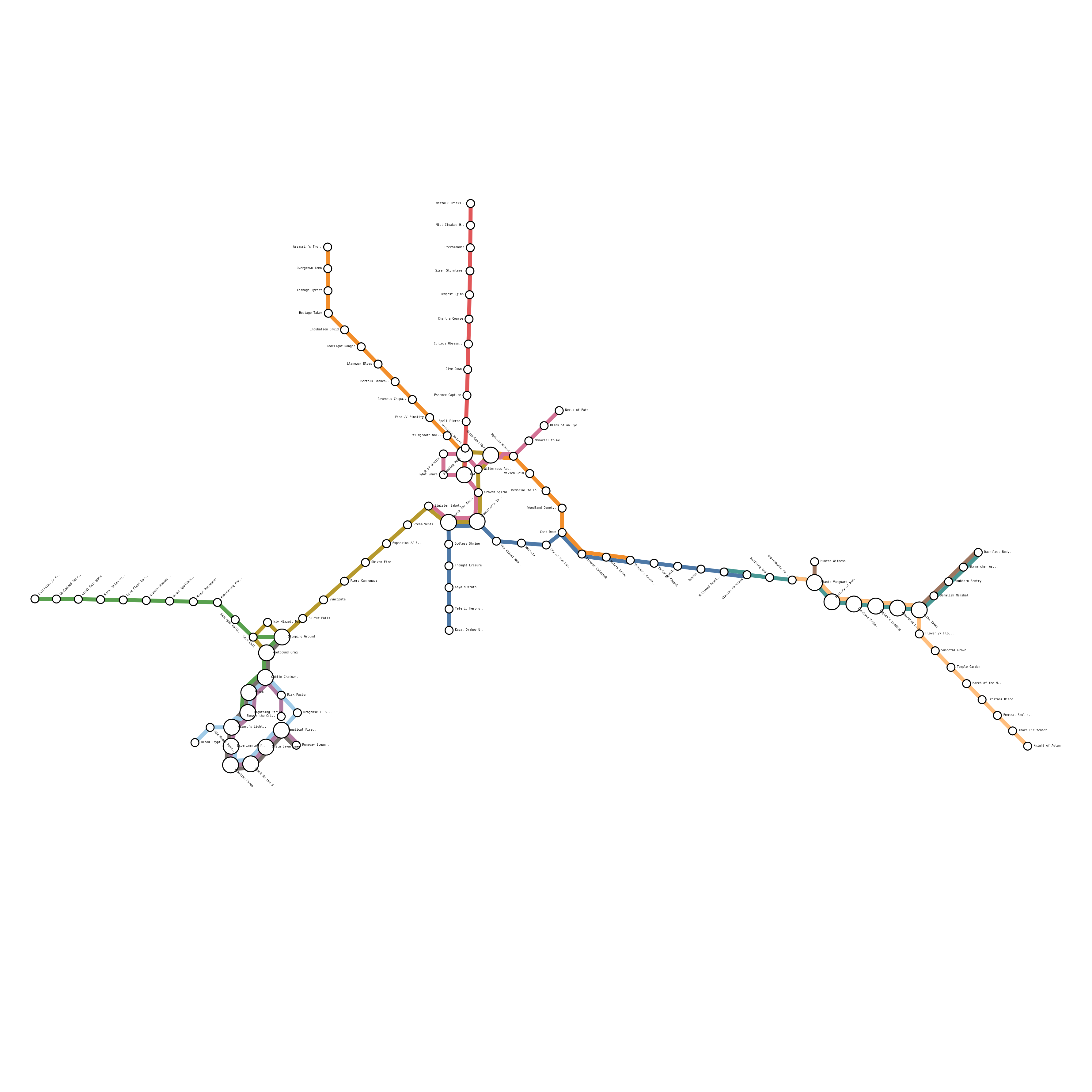}
        \subcaption{This visualization does much better, with a Consecutive Ones score of only 3. The overlapping sets on the righthand side are visualized with simple, easy to read lines running in parallel, instead of complicated, branching cycles.}
    \end{subfigure}
    \end{centering}
    \caption{These two figures illustrate the impact the consecutive ones property has on the final schematization.}
    \label{fig:c1pComparison}
\end{figure*}

\begin{table*}[tb]
 \small
 \begin{tabularx}{\linewidth}{c >{\hsize = 0.45\hsize}X|*{11}{>{\hsize = 1.05\hsize}X}}
 \multicolumn{1}{c}{Factor\rule{0pt}{10ex}} && \begin{rotate}{60}  Gabriel  Score\end{rotate} & \begin{rotate}{60}  Consecutive  Ones\end{rotate} & \begin{rotate}{60}  Monotonicity\end{rotate} & \begin{rotate}{60}  Avg.  Octilinearity\end{rotate} & \begin{rotate}{60}  Max  Octilinearity\end{rotate} & \begin{rotate}{60}  Edge  Crossings\end{rotate} & \begin{rotate}{60}  Self  Crossings\end{rotate} & \begin{rotate}{60}  Line  Crossings\end{rotate} & \begin{rotate}{60}  Running  Time\end{rotate} & \begin{rotate}{60}  Avg.  Edge  Uni.\end{rotate} & \begin{rotate}{60}  Max  Edge  Uni.\end{rotate} \\
 \hline 
 Insert&  p    \newline  $\omega^2$ &   0.00  \newline  0.01 &   0.47  \newline  0.00 & \pg  0.00  \newline  0.02 &   0.22  \newline  0.00 &   0.00  \newline  0.01 &   0.90  \newline  0.00 &   0.00  \newline  0.00 &   0.18  \newline  0.00 &   0.00  \newline  0.00 &   0.00  \newline  0.00 &   0.00  \newline  0.00 \\
 \hline 
 Layout&  p    \newline  $\omega^2$ & \pg  0.00  \newline  0.05 &   0.85  \newline  0.00 & \pg  0.00  \newline  0.04 & \pg  0.00  \newline  0.02 &   0.00  \newline  0.00 & \pg  0.00  \newline  0.04 & \dg  0.00  \newline  0.20 &   0.04  \newline  0.00 & \dg  0.00  \newline  0.19 & \pg  0.00  \newline  0.04 & \pg  0.00  \newline  0.03 \\
 \hline 
 $|\hE|$&  p    \newline  $\omega^2$ & \dg  0.00  \newline  0.37 & \dg  0.00  \newline  0.18 & \dg  0.00  \newline  0.42 & \dg  0.00  \newline  0.22 & \dg  0.00  \newline  0.29 & \dg  0.00  \newline  0.40 & \mg  0.00  \newline  0.08 & \dg  0.00  \newline  0.37 &   0.00  \newline  0.00 & \dg  0.00  \newline  0.47 & \dg  0.00  \newline  0.19 \\
 \hline 
 $|\hV|$&  p    \newline  $\omega^2$ & \dg  0.00  \newline  0.16 & \mg  0.00  \newline  0.06 & \mg  0.00  \newline  0.08 & \pg  0.00  \newline  0.02 & \pg  0.00  \newline  0.03 & \mg  0.00  \newline  0.12 &   0.00  \newline  0.01 & \pg  0.00  \newline  0.01 & \dg  0.00  \newline  0.59 & \mg  0.00  \newline  0.08 & \dg  0.00  \newline  0.23 \\
 \hline 
 Schem.&  p    \newline  $\omega^2$ &   0.58  \newline  0.00 &   0.94  \newline  0.00 &   0.00  \newline  0.00 &   0.00  \newline  0.01 & \dg  0.00  \newline  0.17 &   0.04  \newline  0.00 &   0.84  \newline  0.00 &   0.87  \newline  0.00 & \pg  0.00  \newline  0.01 &   0.00  \newline  0.00 & \mg  0.00  \newline  0.12 \\
 \hline 
 Support&  p    \newline  $\omega^2$ & \pg  0.00  \newline  0.02 &   0.06  \newline  0.00 &   0.00  \newline  0.00 &   0.00  \newline  0.00 &   0.32  \newline  0.00 &   0.00  \newline  0.01 &   0.10  \newline  0.00 &   0.00  \newline  0.01 &   0.00  \newline  0.00 &   0.00  \newline  0.00 &   0.95  \newline  0.00 \\
 \hline 
 Insert,Layout&  p    \newline  $\omega^2$ &   0.68  \newline  0.00 &   0.65  \newline  0.00 &   0.00  \newline  0.00 &   0.35  \newline  0.00 &   0.14  \newline  0.00 &   0.00  \newline  0.00 &   0.00  \newline  0.01 &   0.29  \newline  0.00 &   0.00  \newline  0.00 &   0.09  \newline  0.00 &   0.26  \newline  0.00 \\
 \hline 
 Insert,$|\hE|$&  p    \newline  $\omega^2$ &   0.00  \newline  0.00 &   0.75  \newline  0.00 &   0.00  \newline  0.00 &   0.03  \newline  0.00 &   0.01  \newline  0.00 &   0.97  \newline  0.00 &   0.17  \newline  0.00 &   0.40  \newline  0.00 &   0.00  \newline  0.00 &   0.72  \newline  0.00 &   0.01  \newline  0.00 \\
 \hline 
 Insert,$|\hV|$&  p    \newline  $\omega^2$ &   0.00  \newline  0.00 &   0.95  \newline  0.00 &   0.00  \newline  0.01 &   0.00  \newline  0.00 &   0.02  \newline  0.00 &   0.71  \newline  0.00 &   0.20  \newline  0.00 &   0.89  \newline  0.00 &   0.00  \newline  0.00 &   0.02  \newline  0.00 &   0.02  \newline  0.00 \\
 \hline 
 Insert,Schem.&  p    \newline  $\omega^2$ &   0.70  \newline  0.00 &   0.78  \newline  0.00 &   0.29  \newline  0.00 &   0.04  \newline  0.00 &   0.01  \newline  0.00 &   0.84  \newline  0.00 &   0.14  \newline  0.00 &   0.71  \newline  0.00 &   0.98  \newline  0.00 &   0.00  \newline  0.00 &   0.04  \newline  0.00 \\
 \hline 
 Insert,Support&  p    \newline  $\omega^2$ &   0.44  \newline  0.00 &   0.94  \newline  0.00 &   0.95  \newline  0.00 &   0.39  \newline  0.00 &   0.35  \newline  0.00 &   0.53  \newline  0.00 &   0.65  \newline  0.00 &   0.08  \newline  0.00 &   0.06  \newline  0.00 &   0.45  \newline  0.00 &   0.77  \newline  0.00 \\
 \hline 
 Layout,$|\hE|$&  p    \newline  $\omega^2$ & \pg  0.00  \newline  0.03 &   0.07  \newline  0.00 & \pg  0.00  \newline  0.02 &   0.00  \newline  0.01 &   0.16  \newline  0.00 & \pg  0.00  \newline  0.04 & \mg  0.00  \newline  0.12 &   0.01  \newline  0.00 &   0.00  \newline  0.00 & \pg  0.00  \newline  0.01 & \pg  0.00  \newline  0.01 \\
 \hline 
 Layout,$|\hV|$&  p    \newline  $\omega^2$ & \pg  0.00  \newline  0.01 &   0.01  \newline  0.00 &   0.38  \newline  0.00 &   0.00  \newline  0.00 &   0.02  \newline  0.00 &   0.00  \newline  0.00 & \pg  0.00  \newline  0.01 &   0.36  \newline  0.00 & \dg  0.00  \newline  0.16 &   0.13  \newline  0.00 &   0.03  \newline  0.00 \\
 \hline 
 Layout,Schem.&  p    \newline  $\omega^2$ &   0.68  \newline  0.00 &   0.98  \newline  0.00 &   0.97  \newline  0.00 &   0.00  \newline  0.00 &   0.16  \newline  0.00 &   0.14  \newline  0.00 &   0.75  \newline  0.00 &   0.97  \newline  0.00 &   0.95  \newline  0.00 &   0.00  \newline  0.00 &   0.00  \newline  0.00 \\
 \hline 
 Layout,Support&  p    \newline  $\omega^2$ &   0.00  \newline  0.01 &   0.01  \newline  0.00 &   0.86  \newline  0.00 &   0.08  \newline  0.00 &   0.20  \newline  0.00 &   0.00  \newline  0.00 &   0.06  \newline  0.00 &   0.00  \newline  0.01 &   0.01  \newline  0.00 &   0.01  \newline  0.00 &   0.14  \newline  0.00 \\
 \hline 
 $|\hE|$,$|\hV|$&  p    \newline  $\omega^2$ & \mg  0.00  \newline  0.11 & \dg  0.00  \newline  0.15 & \pg  0.00  \newline  0.04 & \pg  0.00  \newline  0.05 & \pg  0.00  \newline  0.02 & \mg  0.00  \newline  0.09 &   0.00  \newline  0.01 & \mg  0.00  \newline  0.09 &   0.00  \newline  0.00 & \pg  0.00  \newline  0.03 &   0.00  \newline  0.01 \\
 \hline 
 $|\hE|$,Schem.&  p    \newline  $\omega^2$ &   0.94  \newline  0.00 &   1.00  \newline  0.00 &   0.28  \newline  0.00 & \dg  0.00  \newline  0.17 & \pg  0.00  \newline  0.05 &   0.24  \newline  0.00 &   0.97  \newline  0.00 &   1.00  \newline  0.00 &   0.84  \newline  0.00 & \pg  0.00  \newline  0.06 & \pg  0.00  \newline  0.06 \\
 \hline 
 $|\hE|$,Support&  p    \newline  $\omega^2$ &   0.00  \newline  0.01 &   0.00  \newline  0.00 &   0.39  \newline  0.00 &   0.03  \newline  0.00 &   0.08  \newline  0.00 &   0.00  \newline  0.01 &   0.41  \newline  0.00 &   0.22  \newline  0.00 &   0.00  \newline  0.00 &   0.02  \newline  0.00 &   0.84  \newline  0.00 \\
 \hline 
 $|\hV|$,Schem.&  p    \newline  $\omega^2$ &   0.60  \newline  0.00 &   1.00  \newline  0.00 &   0.66  \newline  0.00 &   0.00  \newline  0.01 & \pg  0.00  \newline  0.02 &   0.66  \newline  0.00 &   0.93  \newline  0.00 &   1.00  \newline  0.00 &   0.00  \newline  0.01 & \mg  0.00  \newline  0.07 & \pg  0.00  \newline  0.02 \\
 \hline 
 $|\hV|$,Support&  p    \newline  $\omega^2$ &   0.00  \newline  0.00 &   0.41  \newline  0.00 &   0.00  \newline  0.00 &   0.70  \newline  0.00 &   0.02  \newline  0.00 &   0.05  \newline  0.00 &   0.00  \newline  0.00 &   0.70  \newline  0.00 &   0.00  \newline  0.00 &   0.40  \newline  0.00 &   0.22  \newline  0.00 \\
 \hline 
 Schem.,Support&  p    \newline  $\omega^2$ &   0.00  \newline  0.00 &   0.94  \newline  0.00 &   0.76  \newline  0.00 &   0.04  \newline  0.00 &   0.54  \newline  0.00 &   0.18  \newline  0.00 &   0.49  \newline  0.00 &   0.88  \newline  0.00 &   0.72  \newline  0.00 &   0.02  \newline  0.00 &   0.01  \newline  0.00 \\
 \hline 
 \end{tabularx}
\caption{ Summarizes the impact of different factors on each metric's performance, based on the results of our ANOVA tests. Each column represents a metric while each row represents a factor or pair of factors. The top value in each cell is the p-value of the effect of the factor on the metric, while the bottom value is the $\omega^2$ measure of effect size. Additionally, a cell is colored deep, medium, or light green if that factor has a large ($\omega^2>0.14$), medium ($0.06<\omega^2\leq0.14$) or small ($0.01<\omega^2<0.06$) effect on that metric, respectively. A cell is left white if that factor's effect is very small ($\omega^2\leq0.01$). Pairs of factors represent interaction terms; when a cell in a row corresponding to an interaction term is colored, it means that the value of that metric behaves differently from how you would guess based on looking at the two interacting factors in isolation. All figures are rounded to two decimal places.}
\label{tab:anova-results2}
\end{table*}

\end{document}